\documentclass[onecolumn,12pt,journal,final]{IEEEtran}

\usepackage{amsfonts}
\usepackage[dvips]{graphicx}
\usepackage{times}
\usepackage{color}
\usepackage{cite}
\usepackage{amsmath}
\usepackage{cases}
\usepackage{array}
\usepackage{dsfont}
\usepackage{amssymb}

\usepackage{stfloats}
\usepackage{slashbox}
\usepackage{graphicx}
\usepackage{footnote}
\usepackage{booktabs}
\usepackage{array}
\usepackage{multirow}
\usepackage{bm}
\usepackage{empheq}
\usepackage[labelformat=simple]{subcaption}

\usepackage{amsthm}

\allowdisplaybreaks[1]
\theoremstyle{plain}
\newtheorem{theorem}{Theorem}
\newtheorem{corollary}{Corollary}
\newtheorem{lemma}{Lemma}

\theoremstyle{definition}

\begin{document}

\title{Fundamental Storage-Latency Tradeoff in Cache-Aided MIMO Interference Networks }
\author{\IEEEauthorblockN{Youlong Cao, Meixia Tao, Fan Xu and Kangqi Liu \\}

\thanks{This work is presented in part at the 2016 IEEE Global Telecommunications Conference \cite{long}. The authors are with the Department of Electronic Engineering, Shanghai Jiao Tong University,  Shanghai, China. Emails: \{caoyoulong, mxtao, xxiaof\}@sjtu.edu.cn, k.liu.cn@ieee.org.}
}
\maketitle

\begin{abstract}

Caching is an effective technique to improve user perceived experience for content delivery in wireless networks. Wireless caching differs from traditional web caching in that it can exploit the broadcast nature of wireless medium and hence opportunistically change the network topologies. This paper studies a cache-aided MIMO interference network with $3$ transmitters each equipped with $M$ antennas and $3$ receivers each with $N$ antennas. With caching at both the transmitter and receiver sides, the network is changed to hybrid forms of MIMO broadcast channel, MIMO X channel, and MIMO multicast channels. We analyze the degrees of freedom (DoF) of these new channel models using practical interference management schemes. Based on the collective use of these DoF results, we then obtain an achievable \emph{normalized delivery time} (NDT) of the network, an information-theoretic metric that evaluates the worst-case delivery time at given cache sizes. The obtained NDT is for arbitrary $M$, $N$ and any feasible cache sizes. It is shown to be optimal in certain cases and within a multiplicative gap of $3$ from the optimum in other cases. The extension to the network with arbitrary number of transmitters and receivers is also discussed.
%
%
\end{abstract}


\section{Introduction}
\subsection{Motivation}
Over the last decade, the ever-growing mobile cellular traffic has undergone a fundamental shift from voices and messages to rich content distribution, such as video streaming. In particular, video traffic amounts for more than 50$\%$ of the total mobile data traffic in 2015 and is foreseen to contribute 75$\%$ in 2020 \cite{cisco}. An important feature of video contents is that they are cachable and the same content can be requested by many users. Wireless caching is to prefetch the popular contents at the wireless edge, such as local base stations or mobile users, during the off-peak time in order to reduce the peak data traffic and improve user perceived experience. Caching at the wireless edge can be regarded as an effective way to trade the scarce communication bandwidth with the more sustainable storage size through traffic time shifting. It has attracted significant attention from both academia and industry recently, see for example \cite{caching1,caching2,caching3} and references therein.

Traditional caching has been long proposed in computer networks for reducing the downloading delay \cite{dowdy1982} since a requested file can be obtained in the local cache without resorting to a remote server.  Wireless caching differs from traditional caching in that it can exploit the broadcast nature of wireless medium and hence opportunistically change the network topologies. A fundamental question about wireless caching is what and how much gain it can achieve. This has driven the study of fundamental limits of caching in various wireless systems, including broadcast channel \cite{zhang2015fundamental}, interference networks \cite{Ali2,xu,xu2,niesendof,Ali3}, partially connected networks \cite{Xinping_topo}, device-to-device networks \cite{d2d_1,d2d_twc1,d2d_twc2}, and fog radio access networks \cite{fogRan}.

This work aims to investigate the fundamental limits of caching in wireless MIMO interference networks where each node is equipped with both a local cache and multiple antennas. The system operates in two phases. In the cache placement phase, which usually takes places in a large time scale (e.g. a day or an hour), each node prefetches some file bits from a library into its local cache. In the content delivery phase, which happens in a small time scale (e.g. second), each transmitter sends the messages according to the receiver requests, cache states, and the MIMO channel conditions. Our goal is to characterize the storage-latency tradeoff through the careful design of cache placement and content delivery.

\subsection{Related Works}

The fundamental limits of caching at the receiver side were first studied in \cite{Alilimits} for a shared link with one server and multiple cache-aided receivers. The study in \cite{Alilimits} shows that caching can exploit multicast opportunities even when user demands are different, and hence greatly reduces the traffic load over the shared link. This is enabled by proper file splitting during the cache placement phase and coded transmission during the content delivery phase, known as \emph{coded caching}. The benefits of caching at the transmitter side were studied in \cite{Ali2} for a $3\times 3$ interference channel. It is shown that caching can induce transmitter cooperation and hence allows interference coordination for throughput enhancement.

The limits of caching when equipped at both the transmitter and receiver sides were investigated in \cite{xu,xu2,Ali3,niesendof} very recently, which all considered a general interference network but with different restrictions and performance metrics. The works \cite{xu,xu2} characterized the tradeoff between storage size and content delivery time, in terms of an information-theoretic metric, \emph{normalized delivery time} (NDT). An achievable NDT is obtained in \cite{xu} for an interference network with arbitrary number of transmitters and arbitrary number of receivers at any feasible cache sizes. Their achievable NDT is optimal at certain cache size regions and is within a bounded multiplicative gap to a theoretical lower bound at other cache size regions. The study in \cite{xu} reveals that, with a novel cooperative transmitter and receiver caching strategy, the interference network can be turned opportunistically into more favorable channels, including X channel, broadcast channel, multicast channel, and a hybrid form of these channels. In \cite{niesendof}, an order-optimal approximation on the system performance for arbitrary number of transmitters and receivers was presented. But their analysis is limited to the case where the accumulated cache size at the transmitter side is large enough to cache all the files and only hybrid X-multicast channel is considered. The work \cite{Ali3}, on the other hand, adopted the standard sum degrees of freedom (DoF) to characterize the performance and their analysis is restricted to one-shot linear transmission schemes.
The aforementioned studies on fundamental limits of caching at both transmitter and receiver sides are limited to the single-antenna interference network.


Note that a crucial step in analysing the performance of cache-aided interference networks is to derive the DoF, a capacity approximation at high signal-to-noise ratio (SNR) regime, of the new network topologies formed by caching, for example the X-multicast channel \cite{xu, niesendof}. DoF characterizations for a wide variety of MIMO channels have been considered recently, in particular for MIMO interference channel \cite{2use_ic,Cadambe,wang2014subspace, ic_generalsetting} and MIMO X channel \cite{Maddah,cadambe2009xchannel,sun_xchannel}. However, the DoF results of these MIMO channels with multicast traffic and/or transmitter cooperation remain unsolved in general.

\subsection{Our Contribution}
In this paper, we study a cache-aided MIMO interference network with three transmitters and three receivers, as shown in Fig. \ref{system}. Each transmitter is equipped with $M$ antennas and a local cache of normalized size $\mu_T$, and each receiver with $N$ antennas and a local cache of normalized size $\mu_R$. The performance is characterized by NDT, the same information-theoretic metric applied in \cite{simeone, xu, fogRan}. This work is a non-trivial extension of \cite{xu} due to the deployment of multiple antennas. Preliminary results in the special case with symmetric antenna configuration $M=N$ are presented in the conference paper \cite{long}. This journal paper considers the more general case with arbitrary $M$ and arbitrary $N$. The main contributions and findings of this paper are summarized as follows:

%
%
%

$\bullet$ \emph{An achievable NDT}: We adopt the same cooperative Tx/Rx caching scheme proposed in \cite{xu,xu2} for file placement, but design different and more practical transmission schemes for content delivery. An achievable NDT is obtained by solving a linear programming problem of file splitting or, equivalently, memory sharing coefficients.
%
%
The achievable NDT is for any number of transmit antennas $M$, any number of receive antennas $N$, and any feasible cache size tuple $(\mu_R, \mu_T)$. Its closed-form expression is piecewise-linearly decreasing with the normalized cache sizes, which reflects the caching gain. Each additive item in the expression is inversely proportional to the number of antennas, which reflects the spatial multiplexing gain induced by MIMO. It also found, interestingly, that the traditional equal file splitting strategy \cite{xu,xu2,Ali3,niesendof} is not always optimal at integer points\footnote{The equal file splitting strategy is to split each file into $\binom{3}{3\mu_R}$$\binom{3}{3\mu_T}$ equal-sized subfiles, each cached in $3\mu_R$ receivers and $3\mu_T$ transmitters when $\mu_R,\mu_T\in\big\{0,\frac{1}{3},\frac{2}{3},1\big\}$. Due to that $3\mu_R=m$ and $3\mu_T=n$ with $m,n$ being integers, these points are called integer points\cite{xu} or corner points \cite{Alilimits}.}.

$\bullet$ \emph{DoF of new MIMO channel models}: A crucial step in analyzing the achievable NDT is to derive the DoF of the new network topologies formed by different file placement patterns during the content delivery phase. In this work, several new channel models are formed, including $3\times 3$ partially cooperative MIMO X channel, $3\times 3$ MIMO X-multicast channel, and $3\times 3$ partially or fully cooperative MIMO X-multicast channel. We derive the achievable DoF per user of these channels by using linear precoding based interference management schemes with finite symbol extensions such as interference alignment, interference neutralization and zero forcing. We would like to remark that a related but different effort is the study of DoF region of MIMO interference network with general message demands in \cite{ic_generalsetting}. Our channel models differ from \cite{ic_generalsetting} in that (1) each transmitter has multiple messages to send and can cooperate with each other; (2) the antenna configurations at the transmitter and receiver sides are asymmetric. Another related effort is the study of DoF region of X channel with multicast traffic in \cite{Wang_it16}, which, however only considers single antenna.

$\bullet$ \emph{A lower bound of the minimum NDT}: We also obtain a theoretical lower bound of the minimum NDT of the considered cache-aided MIMO interference network by using a cut-set like argument. This lower bound has no restriction on the linearity of MIMO transmission schemes and allows arbitrary intra-file coding but not inter-file coding at the cache placement phase. With this lower bound, we show that our achievable upper bound is optimal for certain antenna configurations and cache size regions. Analysis also shows that the maximum multiplicative gap between the upper and lower bounds is $3$.

%

Notations: $x$, ${\bf x}$, ${\bf X}$ and $\cal{X}$ denotes scalar, vector, matrix and set, respectively. $\Theta(x)$ denotes that $\lim\limits_{x\rightarrow\infty}\frac{\Theta(x)}{x}=1$. $(\cdot)^{T}$ denotes the transpose of a matrix . tr$(\bf X)$ and ${\mbox{null} ({\bf X})}$ stand for the trace and the null space of the matrix ${\bf X}$. $[n]$ denotes the set $\{1,2,\cdots,n\}$ where $n$ is an integer.

\section{System Model and Performance Metrics}
\begin{figure*}[t]
 \vspace{-0.35cm}
\begin{centering}
\includegraphics[scale=0.4]{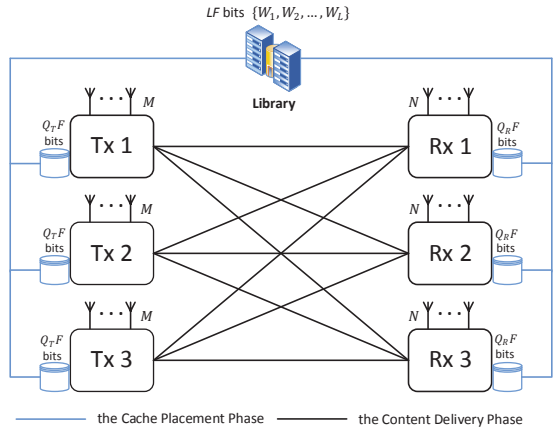}
 \caption{The $3\times 3$ cache-aided MIMO interference network.}\label{system}
\end{centering}
\end{figure*}
We consider a $3\times 3$ cache-aided MIMO interference network as illustrated in Fig. \ref{system}, where each transmitter is equipped with $M$ antennas and each receiver is equipped with $N$ antennas. Each node has a local cache of finite size. Consider a library consisting of $L$ files, denoted by $\{W_1, W_2,\ldots, W_L\}$. Throughout this study, we focus on the case where the number of files $L$ is larger than or equal to the number of receivers, i.e., $L\geq3$. Each file has the same length of $F$ bits. Each transmitter can cache $Q_{T}F$ bits and each receiver can cache $Q_{R}F$ bits, where $Q_T,Q_R \leq L$. The normalized cache sizes at the transmitter and receiver sides are defined, respectively, as
\begin{equation}
\mu_{T}=\frac{Q_{T}}{L}, \text{and }\mu_{R}=\frac{Q_{R}}{L}.
\end{equation}
This work focuses on the feasible cache size region \cite{xu} \cite{xu2}:
\begin{equation}\label{uu2}
\left \{\begin{split}
&  0\leq\mu_R,\mu_T\leq 1, \\
& \mu_R+3\mu_T \geq 1.
\end{split}
\right.
\end{equation}

The communication involves two phases, the cache placement phase, which takes place in a large time scale, and the content delivery phase, which happens in a small time scale. During the cache placement phase, each transmitter $i$ has a caching function
\begin{equation}
\phi_{i}: [2^{F}]^L  \rightarrow [2^ {\lfloor FQ_{T} \rfloor}],
\end{equation}
mapping the $L$ files in the library to its local cache content $U_{i}\triangleq \phi_{i}(W_{1},W_{2},\ldots,W_{L})$, for $i\in [3]$. Each receiver $j$ also has a caching function
\begin{equation}
\psi_{j}: [2^{F}]^L  \rightarrow [2^ {\lfloor FQ_{R} \rfloor}],
\end{equation}
mapping the $L$ files to its local cache content $V_{j} \triangleq \psi_{j}(W_{1},W_{2},\ldots,W_{L})$, for $j\in [3]$. As in \cite{xu,simeone}, it is assumed that the caching functions $\{\phi_{i}, \psi_{j}\}$ allow arbitrary intra-file coding, but do not allow inter-file coding.

 In the content delivery phase, each receiver $j$ requests a file $W_{d_j}$ from the library, where $d_j \in [L]$. We denote ${\bf d}  \triangleq [d_1,d_2,d_3]^{T}$ as the demand vector. Each transmitter further consists of an encoding function
\begin{equation}
\Lambda_{i}:[2^ {\lfloor FQ_{T} \rfloor}]\times[L]^{3}\times\mathbb{C}^{3N\times 3M}\rightarrow\mathbb{C}^{M\times T},
\end{equation}
where $T$ is the block length of the code and depends on the receiver demand ${\bf d}$ and the network channel state information (CSI) ${\bf H} = \{ {\bf H}_{i,j} \in\mathbb{C}^{N \times M}: {i\in[3]}, {j\in[3]} \}$. Each ${\bf H}_{i,j}$ is the channel matrix from each transmitter $i$ to each receiver $j$, whose entries are drawn independently and identically distributed (i.i.d) from a continuous distribution, and remain invariant within each codeword transmission. Transmitter $i$ uses $\Lambda_{i}$ to map its local cache content $U_{i}$, receiver demands ${\bf d}$ and the network CSI ${\bf H}$ to the signal vectors $\big[{\bf x}_{i}(t)\big]_{t=1}^{T}\triangleq\Lambda_{i}(U_{i},{\bf d},{\bf H})$, which is subject to a power constraint $\text{tr}\left[{\bf x}_{i}(t){\bf x}^H_{i}(t)\right]\leq P$.
In each time slot $t\in [T]$, the received signal at each receiver $j$, denoted as ${\bf y}_j(t)\in\mathbb{C}^{N \times 1}$, can be expressed as
\begin{equation}
{\bf y}_{j}(t)=\sum\limits_{i=1}^{3}{\bf H}_{ij}{\bf x}_{i}(t)+{\bf n}_{j}(t),\quad \forall j \in [3],
\end{equation}
where ${\bf n}_{j}(t)$ denotes the additive white Gaussian noise (AWGN) vector at receiver $j$, with each element being independent and having zero mean and unit variance. In this paper, we assume that the network CSI is available at all transmitters and receivers.
The decoding function $\Gamma_j$ at receiver $j$ can be defined as:
\begin{equation}
\Gamma_{j}:[2^ {\lfloor FQ_{R} \rfloor}]\times\mathbb{C}^{N\times T}\times\mathbb{C}^{3N\times 3M}\times[L]^{3} \rightarrow[2^ {\lfloor F \rfloor}].
\end{equation}
Each receiver $j$ uses $\Gamma_{j}$ to estimate ${\hat W
}_{j}\triangleq \Gamma_{j}(V_{j},\big[{\bf y}_{j}(t)\big]_{t=1}^{T},{\bf H},{\bf d})$ of its desired file ${W_{d_j}}$, with its cached content $V_{j}$ and the channel realization ${\bf H}$. The worst-case error probability is
\begin{equation}
\max\limits_{{\bf d}\in[L]^{3}}\max\limits_{j\in[3]} \mathbb{P}({\hat W
}_{j}\neq W_{d_j}).
\end{equation}
The given caching and coding functions $\{\phi_{i},\Lambda_{i},\psi_{j},\Gamma_{j}\}$ are said to be feasible if, for almost all channel realizations ${\bf H}$, the worst-case error probability approaches 0 when $F\rightarrow \infty$.

In this work, we adopt the following performance metric to characterize the fundamental storage-latency tradeoff\footnote{The performance metric NDT is first proposed in \cite{simeone} for wireless networks with transmitter caches only. It is then scaled by the number of receivers and renamed as fractional delivery time (FDT) by taking receiver caches into account in \cite{xu,xu2,long} as well as the prior version of this paper. During the paper revision, we have removed the scaling and changed back to NDT for consistency with \cite{simeone} as suggested by reviewers. }.

\textbf{Definition 1 \cite{simeone}: }
For any given feasible caching and coding scheme at given normalized cache sizes $\mu_T$ and $\mu_R$, the \emph{normalized delivery time} (NDT) is defined as
\begin{equation}
\tau(\mu_{R},\mu_{T})\triangleq\lim\limits_{P \to \infty}\lim\limits_{F \to \infty}\sup\frac{\max\limits_{\bf d}{T}}{F/\log P}.
\end{equation}
The minimum NDT is defined as
\begin{equation}
\tau^{\ast}(\mu_{R},\mu_{T})=\inf\{\tau(\mu_{R},\mu_{T}): \tau(\mu_{R},\mu_{T})\text{ is achievable.}\}
\end{equation}

Note that $F/\log P$ is the delivery time of transmitting one file of $F$ bits over a point-to-point Gaussian channel with single antenna in the high signal-to-noise ratio (SNR) regime. An NDT of $\tau$ thus indicates that the worst-case time required to serve any possible demand vector $\textbf{d}$ is $\tau$ times of this reference time period.

\textbf{Remark 1 \cite{xu}:}
Let $R$ denote the worst-case traffic load per user with respect to the file size $F$, and $d\cdot\log{P} + o(\log{P})$ denote the per-user capacity of the network in the high SNR regime at a given caching and coding scheme.  By Definition 1, NDT can be approximately expressed as
\begin{equation}\label{eqn NDTtau}
\tau=\frac{RF/{(d\cdot \log P)}} {F/\log P} = \frac{R}{d}.
\end{equation}
Thus, NDT characterizes the asymptotic delivery time of the actual traffic load per user, $R$, at a transmission rate specified by the DoF per user, $d$, when both transmit power $P$ and file size $F$ go to infinity.

%

\begin{figure*}[t]
\begin{centering}
\includegraphics[scale=0.4]{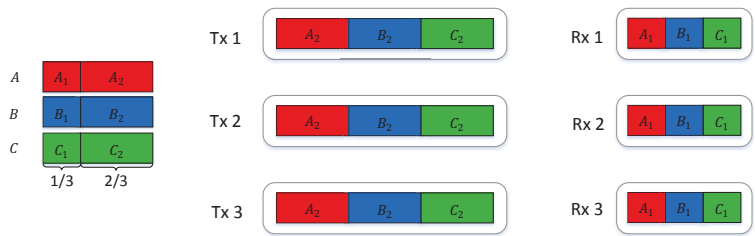}
 \caption{File splitting and cache placement at $\mu_R=\frac{1}{3}$, $\mu_T=\frac{2}{3}$ with $M=N$. } \label{example}
\end{centering}
\end{figure*}

\textbf{Example 1.} Consider a $3\times 3$ MIMO interference network with the normalized cache sizes $\mu_R=\frac{1}{3}$ and $\mu_T=\frac{2}{3}$ under symmetric antenna setting $M=N$.  The cache placement strategy is shown in Fig. \ref{example}, where each file is split into two subfiles, one with $\frac{1}{3}F$ bits and cached in all receivers, the other with $\frac{2}{3}F$ bits and cached in all transmitters. During the delivery phase, consider the worst case where the three receivers request distinct files, denoted as $A$, $B$, $C$, respectively. Then each receiver only needs the subfile with the length of $\frac{2}{3}F$ bits that it does not cache, which is available at all the three transmitters. The traffic load per user is $R=\frac{2}{3}$. The network topology can be viewed as a virtual MIMO broadcast channel where the virtual transmitter has $3M$ antennas and each receiver has $M$ antennas. The DoF per user of this channel is $d=M$. By Remark 1, the achievable NDT is $\tau=\frac{2}{3M}$.

\section{Main results}
In this section, we present our main findings on the minimum NDT in the $3\times 3$ cache-aided MIMO interference network.

\begin{theorem}[Upper Bound]\label{thm upper}
 Consider the $3\times 3$ cache-aided MIMO interference network where each transmitter is equipped with $M$ antennas and a cache of normalized size $\mu_{T}$, and each receiver is equipped with $N$ antennas and a cache of normalized size $\mu_{R}$. An achievable NDT based on linear transmission schemes with finite symbol extensions is given by $\tau_{u}$, the optimal solution of the following linear programming problem:
\begin{subequations}
\begin{align}
\mathcal{P}_1:\quad\tau_{u}\triangleq &  \min \limits_{\{ \beta _{mn}\}} \quad \sum\limits_{(m,n)\in \cal{A}} \beta_{mn}\frac{1-m/3}{d_{mn}} \\
  &\ \text{s.t. } \quad \ \sum\limits_{(m,n)\in \cal{A}}  \beta_{mn}=1, \label{b1}\\
  &\quad \quad  \quad \; \sum\limits_{(m,n)\in \cal{A}} \beta_{mn}{\bm \mu}_{mn}^{o}\leq{\bm \mu},\label{burut}\\
  &\quad \quad  \quad \quad 0\leq \beta_{mn}\leq1,\quad \forall (m,n)\in \cal{A},
\end{align}
\end{subequations}
where ${\cal A}=\{(m,n):  m+3n\geq3, m,n\in \{0,1,2,3\}\}$; ${\bm \mu}=[\mu_R,\mu_T]^T$ denotes any feasible point in the cache size region; ${\bm \mu}_{mn}^{o}=[\frac{m}{3},\frac{n}{3}]^T$ denotes the integer point with $(\mu_R=\frac{m}{3},\mu_T=\frac{n}{3})$ in the cache size region; $\beta_{mn}$ is the (memory sharing) parameter to be optimized; and $d_{mn}$ is given below:
\begin{equation*}
\begin{aligned}
& \ d_{01}= \min \left\{\frac{k^-}{2-\frac{1}{\xi}}, \frac{k^+}{2+\frac{1}{\xi}}\right\}, \quad\quad \quad d_{03}=\min\{M,N\},\quad \quad \quad d_{13}=\min\{N,2M\},\\
&\begin{array}{ccc}
  d_{02}=\begin{cases}

                                  N,  & \frac{N}{M} \in \left(0,\frac{2}{3}\right] \\
                                  \frac{2M}{3}, & \frac{N}{M} \in \left(\frac{2}{3},\frac{5}{3}\right] \\
                                  \frac{2N}{5}, & \frac{N}{M} \in \left(\frac{5}{3},\frac{5}{2}\right] \\
                                  M,  & \frac{N}{M} \in \left(\frac{5}{2},\infty \right) \\
\end{cases}, & d_{11}=
             \begin{cases}
             \frac{6N}{7}, & \frac{N}{M}\in\left(0,1\right]\\
             \frac{6M}{7},  & \frac{N}{M}\in\left(1,\frac{9}{7}\right] \\
             \frac{2N}{3},  & \frac{N}{M}\in\left(\frac{9}{7},3\right] \\
             2M,   & \frac{N}{M} \in\left(3,\infty\right) \\
\end{cases}, & d_{12}=
             \begin{cases}
             N,   &\frac{N}{M}\in\left(0,1\right]\\
             M,   &\frac{N}{M}\in\left(1,\frac{3}{2}\right] \\
             \frac{2N}{3},  &\frac{N}{M}\in\left(\frac{3}{2},3\right] \\
             2M,   &\frac{N}{M}\in\left(3,\infty\right) \\
\end{cases},
\end{array} \\
& d_{21}=d_{22}=d_{23}=\min\{N,3M\},
\end{aligned}
\end{equation*}
where $k^-=\min\{M,N\}$, $k^+=\max\{M,N\}$ and $\xi=\lceil\frac{k^-}{k^+-k^-}\rceil$.
\end{theorem}

\textbf{Remark 2:}
The linear programming (LP) problem $\mathcal{P}_1$ in Theorem \ref{thm upper} can be solved efficiently. The explicit and closed-form, but somewhat tedious expression of $\tau_u$ is given in Appendix A. It is seen from Appendix A that the achievable NDT decreases piecewise linearly with the normalized cache sizes and each additive item of NDT is, in general, inversely proportional to the number of antennas. The latter property explicitly shows the multiplexing gain induced by MIMO. Moreover, the antenna configuration (i.e., the ratio $N/M$) determines the partition of the cache size region.

\textbf{Remark 3:}
In the special case with symmetric antenna configuration, i.e., $M=N$, the achievable NDT reduces to the results in \cite{long}. Furthermore, when $M=N=1$, the obtained NDT is numerically at most $1.2$ times of the one in the single antenna case \cite{xu2}. The slight increase in the achievable NDT is due to that we only use linear precoding based interference management schemes with finite symbol extensions.

\begin{theorem}[Lower Bound]\label{thm lower}
 Consider the $3\times 3$ cache-aided MIMO interference network where each transmitter is equipped with $M$ antennas and a cache of normalized size $\mu_{T}$, and each receiver is equipped with $N$ antennas and a cache of normalized size $\mu_{R}$. The minimum NDT is lower bounded by
\begin{equation}
\tau^{\ast} \geq \tau_{l}\triangleq \max\left\{\frac{1}{N}(1-\mu_R),\max\limits_{s\in[3]}\frac{s}{3M}(1-s\mu_R)\right\}.
\end{equation}
\end{theorem}

\textbf{Remark 4:}
By comparing the closed-form upper bound in Appendix A and the lower bound in Theorem \ref{thm lower}, it can be seen that the achievable NDT is optimal under the following conditions:
\begin{enumerate}
  \item $\frac{N}{M}\in\left(0,\frac{1}{3}\right]$ and $(\mu_{R},\mu_{T})\in\{(\mu_{R},\mu_{T}):\mu_{R}+3\mu_{T}\geq 1,\mu_{R}\leq 1,\mu_{T}\leq 1\}$;
  \item $\frac{N}{M}\in\left(0,1\right]$ and $(\mu_{R},\mu_{T})\in\{(\mu_{R},\mu_{T}):\mu_{R}+\mu_{T}\geq 1,\mu_{R}\leq 1,\mu_{T}\leq 1\}$;
  \item $\frac{N}{M}\in\left(0,2\right]$ and $(\mu_{R},\mu_{T})\in\{(\mu_{R},\mu_{T}):\mu_{R}+\mu_{T}\geq 1,2\mu_{R}+\mu_{T}\geq \frac{5}{3},\mu_{R}\leq 1,\mu_{T}\leq 1\}$;
  \item $\frac{N}{M}\in\left(0,\infty\right)$ and $(\mu_{R},\mu_{T})\in\{(\mu_{R},\mu_{T}):\mu_{R}+\mu_{T}\geq 1,\frac{2}{3}\leq\mu_{R}\leq 1,\mu_{T}\leq 1\}$.
\end{enumerate}
%
%
%

\begin{corollary}[Multiplicative Gap]\label{corollary multiplicative gap}
The multiplicative gap between the upper and lower bounds of the minimum NDT for the considered network is at most $3$.
\end{corollary}

The proof of Theorem \ref{thm upper} will be given in Sections IV and V. The proofs of Theorem \ref{thm lower} and Corollary \ref{corollary multiplicative gap}  will be given in Section VI.

\section{Caching and delivery Scheme}
The achievable upper bound of minimum NDT in Theorem \ref{thm upper} is based on the same cache placement strategy in \cite{xu} but with different delivery scheme due to the deployment of multiple antennas. In this section, we first review the file splitting and caching strategy proposed in \cite{xu} for self-completeness. Then we present the delivery scheme in detail.

Since each transmitter and receiver can decide whether to cache each bit of each file, there are $2^{6}=64$ possible cache states for each bit. Not every cache state is, however, legitimate, due to that every bit of the file which is not cached simultaneously in all receivers must be cached in at least one transmitter. This results in a total of 57 legitimate cache states for each bit and the feasible domain of $\mu_R$ and $\mu_T$, given in \eqref{uu2}. We now split each file into 57 subfiles \footnote{Note that  57 is the maximum number of legitimate subfiles by exhausting all the possible combinations. The actual number of subfiles can be much less after optimization since not all possible combinations are needed.}, each corresponding to a unique cache state and having a possibly different length to be optimized.
Define transmitter subset $ {\cal I} \subseteq [3]$ and receiver subset ${\cal J} \subseteq [3]$. Then, let $W_{\kappa r_{\cal{J}}t_{\cal{I}}}$ denote the subfile split from $W_{\kappa}$ that is cached in receiver subset $\cal{J}$ and transmitter subset $\cal{I}$. For example, $W_{\kappa r_{\o}t_{123}}$ is the subfile cached in none of the three receivers but in all three transmitters and $W_{\kappa r_{12}t_{12}}$ is the subfile cached in receiver 1, 2 and transmitters 1, 2. Similarly, we denote $W_{\kappa t_{\cal{I}}}$ as the collection of all subfiles cached in $\cal{I}$, i.e. $W_{\kappa t_{\cal{I}}}=\bigcup\limits_{\cal{J}} W_{\kappa r_{\cal{J}} t_{\cal{I}}}$. As in \cite{xu}, the sizes of the subfiles with the same cardinality of transmitter and receiver subsets are assumed to be equal. Let $m=|{\cal J}|$ and $n=|{\cal I}|$ denote the cardinalities of ${\cal J}$ and ${\cal I}$ respectively, and define $a_{mn}$ as the file splitting ratio to be optimized. Then each subfile $W_{\kappa r_{\cal{J}} t_{\cal{I}}}$ contains $Fa_{mn}$ bits. The splitting ratios must satisfy the following constraints:
\begin{align}
&a_{30}+3a_{31}+3a_{32}+a_{33}+9a_{21}+9a_{22}+3a_{23}+9a_{11}+9a_{12}+3a_{13}+3a_{01}+3a_{02}+a_{03}=1, \label{1}\\
&a_{30}+3a_{31}+3a_{32}+a_{33}+6a_{21}+6a_{22}+2a_{23}+3a_{11}+3a_{12}+a_{13}\leq\mu_{R}, \label{ur} \\
&a_{31}+2a_{32}+a_{33}+3a_{21}+6a_{22}+3a_{23}+3a_{11}+6a_{12}+3a_{13}+a_{01}+2a_{02}+a_{03}\leq\mu_{T}.\label{ut}
\end{align}
Constraint \eqref{1} comes from the total file size  limit, where the multiplier of each splitting ratio $a_{mn}$ indicates the total number of subfiles that have the same length $a_{mn}$. Constraints \eqref{ur} and \eqref{ut} come from the cache size limit in receiver and transmitter, respectively, where the multiplier of each splitting ratio $a_{mn}$ indicates the total number of subfiles that are stored in a same receiver or transmitter, and have the same length $a_{mn}$.

In the delivery phase, we consider the worst-case scenario where all user demands are distinct. When user demands are not all distinct, the same delivery strategy can be applied either directly or by treating the demands as being different. Now, without loss of generality, we assume that receiver 1, 2, 3 request file $W_1$, $W_2$, and $W_3$ respectively.

Similar to \cite{xu}, we first divide all the subfiles to be transmitted into groups according to the number of transmitters and receivers where they are cached, then deliver each group separately. However, the specific delivery strategy for each group is significantly different from \cite{xu} due to the deployment of multiple antennas. Namely, as the crucial step in analyzing the achievable NDT, the DoF analysis of the new channel models formed by the different subfile groups in this work is for multiple antennas, while the DoF analysis in \cite{xu} is for single antenna.

\subsection{Delivery of Subfiles Cached in Zero Receiver and One Transmitter}
\begin{figure*}[t]
\begin{centering}
\includegraphics[scale=0.6]{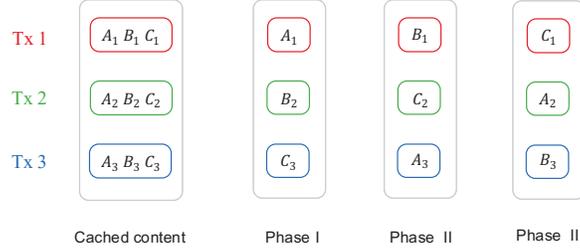}
 \caption{The delivery of subfiles $\{W_{\kappa r_{\o}t_{p}}:\kappa, p\in[3]\}$. Here, we use $A$, $B$ and $C$ to denote the $W_1$, $W_2$ and $W_3$, respectively. Each file is divided into three subfiles of equal size, e.g., $A=(A_{1}, A_{2},A_{3})$, with the subscript indicating at which transmitter this subfiles is to be cached.} \label{a01}
\end{centering}
\end{figure*}
Consider the delivery of subfiles $\{W_{\kappa r_{\o}t_{p}}:\kappa, p\in[3]\}$, each of which is cached at one transmitter but none of the receivers and has fractional length $a_{01}$. The network topology in this case can be seen as a $3\times 3$ MIMO X channel. Previous study on the DoF of MIMO X channel can be found in \cite{cadambe2009xchannel}, but the results require infinite symbol extensions which limits its practical use \cite{jafar1}. In this work, we treat the MIMO X channel as a MIMO interference channel instead, whose optimal DoF is obtained in \cite{wang2014subspace,Davidfeasibility} and only requires linear transmission scheme with finite symbol extensions. The conversion from MIMO X channel to MIMO interference channel is shown in Fig. \ref{a01}, where three phases are needed to deliver the subfiles, and in each phase each transmitter sends an independent message to a different receiver. The DoF per user of the \textbf{$3\times 3$ MIMO interference channel}, denoted as $d_{01}$, is \cite{wang2014subspace,Davidfeasibility} \footnote{In case $d_{01}$ is not an integer, the achievable scheme needs $t$-symbol extension such that $td_{01}$ is an integer.},
\begin{equation}
d_{01}= \min \left\{\frac{k^-}{2-1/\xi}, \frac{k^+}{2+1/\xi}\right\},
\end{equation}
where $k^-=\min\{M,N\}$, $k^+=\max\{M,N\}$ and $\xi=\lceil\frac{k^-}{k^+-k^-}\rceil$. Given that the total amount of bits per user to deliver in each phase is $a_{01}F$ bits, by \eqref{eqn NDTtau}, the NDT over the three-phase transmission can be computed as $\tau=\frac{3a_{01}}{d_{01}}$.

\subsection{Delivery of Subfiles Cached in Zero Receiver and Two Transmitters}
 \begin{figure*}[t]
\begin{centering}
\includegraphics[scale=0.55]{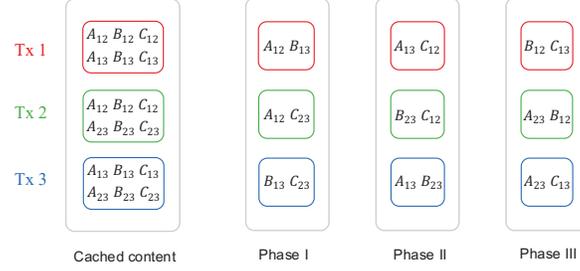}
 \caption{The delivery of subfiles $\{W_{\kappa r_{\o}t_{pq}}: \kappa,p,q\in[3],\ p<q\}$. Each file is divided into three subfiles of equal size, e.g., $A=(A_{12}, A_{13},A_{23})$, with the subscript indicating at which transmitters this subfiles is to be cached.   }\label{a02}
\end{centering}
\end{figure*}
Consider the delivery of subfiles $\{W_{\kappa r_{\o}t_{pq}}: \kappa,p,q\in[3],\ p<q\}$, each of which is cached in two transmitters and none of receivers and has fractional length $a_{02}$. The network topology in this case can be viewed as a \textbf{$3\times 3$ partially cooperative MIMO X channel}, where every set of two transmitters forms a transmit cooperation group and has an independent message to send to each receiver.
\begin{lemma} \label{cooperativeX}
For the $3\times 3$ partially cooperative MIMO X channel, the achievable DoF per user, denoted as $d_{02}$, is given in \eqref{dofa02}.
\begin{equation}\label{dofa02}
d_{02}=\begin{cases}

                                  N,  & \frac{N}{M} \in \left(0,\frac{2}{3}\right] \\
                                  \frac{2M}{3}, & \frac{N}{M} \in \left(\frac{2}{3},\frac{5}{3}\right] \\
                                  \frac{2N}{5}, & \frac{N}{M} \in \left(\frac{5}{3},\frac{5}{2}\right] \\
                                  M,  & \frac{N}{M} \in \left(\frac{5}{2},\infty \right) \\
\end{cases}
\end{equation}
\end{lemma}

\begin{IEEEproof}
The achievable scheme takes three phases, as shown in Fig. \ref{a02}. In each phase, each of the three transmit cooperation groups ($\{1,2\}$, $\{1,3\}$ and $\{2,3\}$) sends one independent message intended to a different receiver. The two interference signals seen by each receiver are cancelled by interference neutralization with linear transmit and receive processing. The detailed proof is given in Appendix B.
\end{IEEEproof}

Based on Lemma~\ref{cooperativeX}, the NDT of these subfiles is $\tau=\frac{3a_{02}}{d_{02}}$.

\subsection{Delivery of Subfiles Cached in Zero Receiver and Three Transmitters}
Consider the delivery of subfiles $\{W_{\kappa r_{\o}t_{123}}:\kappa\in[3]\}$, each of which has fractional length $a_{03}$. Since each subfile is cached in all the three transmitters, the transmitters can fully cooperate. The delivery in this case can be regarded as an \textbf{MIMO broadcast channel} where the virtual transmitter has $3M$ antennas, and each receiver has $N$ antennas. The optimal DoF per user of this channel is $d_{03}=\min\{M,N\}$ \cite{dofbroadcast}. Therefore, the NDT of these subfiles is $\tau=\frac{a_{03}}{d_{03}}$. The delivery scheme in Example $1$ shown in Section II belongs to this case where $a_{03}=\frac{2}{3}$.

\subsection{Delivery of Subfiles Cached in One Receiver and One Transmitter}
Consider the delivery of subfiles $\{W_{\kappa r_{k}t_{p}}: \kappa,k,p\in[3],\  k\neq \kappa\}$, each of which has fractional length $a_{11}$. Since each subfile desired by one receiver is already cached in one of the other receivers, we can use coded multicasting in the delivery phase. For example, transmitter 1 can generate message $W^{\oplus}_{\kappa_{jk}t_1}  \triangleq W_{jr_{k}t_{1}}\oplus W_{kr_{j}t_{1}}$ desired by receiver $j$ and $k$, where $\oplus$ denotes the bit-wise XOR. Now each XORed message is desired by two receivers. The network topology of sending coded subfiles $\{ W^{\oplus}_{r_{jk}t_p}: j,k,p \in[3],\  j<k \}$ becomes a \textbf{$3\times 3$ MIMO X-multicast channel}, where every set of two receivers forms a receive multicast group and each transmitter has an independent message to send to each receive multicast group.

\begin{lemma}\label{X-multicast}
For the $3\times 3$ MIMO X-multicast channel, the achievable DoF per user, denoted as $d_{11}$, is given in \eqref{dofa11}.
\begin{equation}\label{dofa11}
d_{11}=
             \begin{cases}
             \frac{6N}{7}, & \frac{N}{M}\in\left(0,1\right]\\
             \frac{6M}{7},  & \frac{N}{M}\in\left(1,\frac{9}{7}\right] \\
             \frac{2N}{3},  & \frac{N}{M}\in\left(\frac{9}{7},3\right] \\
             2M,   & \frac{N}{M} \in\left(3,\infty\right) \\
\end{cases}
\end{equation}
\end{lemma}

\begin{IEEEproof}
 When the antenna configuration satisfies $N\leq\frac{9}{7}M$, we use linear interference alignment technique so that all the interference signals at each receiver can be aligned in the same direction. When the antenna configuration satisfies $N>\frac{9}{7}M$, the delivery includes three phases and in each phase, each transmitter sends one independent message to a different receive multicast group. The receiver combining matrices are designed to cancel the interference signals. The detailed proof is given in Appendix C.
\end{IEEEproof}
Based on Lemma~\ref{X-multicast}, the NDT of these subfiles is $\tau=\frac{6a_{11}}{d_{11}}$.

\subsection{Delivery of Subfiles Cached in One Receiver and Two Transmitters}
 Similar to Subsection D, coded multicasting gain can be exploited in the delivery of subfiles $\{W_{\kappa r_{k}t_{pq}}: \kappa,k,p,q\in[3],\  k\neq \kappa,\ p<q\}$, each of which has fractional length $a_{12}$. The difference is that each subfile is available at two transmitters and hence transmitter cooperation gain can be exploited. For example, transmitter 1 and transmitter 2 can generate a coded message $W^{\oplus}_{\kappa_{jk}t_{12}} \triangleq W_{jr_{k}t_{12}}\oplus W_{kr_{j}t_{12}}$. The delivery of coded subfiles $\{ W^{\oplus}_{r_{jk}t_{pq}}: j,k,p,q \in[3],\  j<k,\  p<q\}$ can be viewed as a \textbf{$3\times 3$ partially cooperative MIMO  X-multicast channel}, where every set of two receivers forms a receive multicast group, every set of two transmitters forms a transmit cooperation group, and each transmit cooperation group has an independent message for each receive multicast group.
\begin{lemma}\label{pcoperativeX-multicast}
For the $3\times 3$ partially cooperative MIMO X-multicast channel, the achievable DoF per user, denoted as $d_{12}$, is given in \eqref{dofa12}.
\begin{equation}\label{dofa12}
d_{12}=
             \begin{cases}
             N,   &\frac{N}{M}\in\left(0,1\right]\\
             M,   &\frac{N}{M}\in\left(1,\frac{3}{2}\right] \\
             \frac{2N}{3},  &\frac{N}{M}\in\left(\frac{3}{2},3\right] \\
             2M,   &\frac{N}{M}\in\left(3,\infty\right) \\
\end{cases}
\end{equation}
\end{lemma}

\begin{IEEEproof} When the antenna configuration satisfies $N\leq\frac{3}{2}M$, we use the linear interference naturalization by designing the precoding matrices of each transmit cooperation group. When $N>\frac{3}{2}M$, the achievable scheme takes three phases and in each phase, each transmit cooperation group sends one independent message to a different receive multicast group. Each receiver applies zero-forcing processing for interference cancellation. The detailed proof is given in Appendix D.
\end{IEEEproof}
Based on Lemma~\ref{pcoperativeX-multicast}, the NDT of these subfiles is $\tau=\frac{6a_{12}}{d_{12}}$.

\subsection{Delivery of Subfiles Cached in One Receiver and Three Transmitters}
Similar to Subsections D and E, the coded multicasting scheme can also be exploited in the delivery of subfiles $\{W_{\kappa r_{k}t_{123}}: \kappa,k\in[3],\  k\neq \kappa\}$. The difference is that each subfile is available at all the transmitters. For example, all the transmitter can generate message $W^{\oplus}_{\kappa_{jk}t_{123}}  \triangleq W_{jr_{k}t_{123}}\oplus W_{kr_{j}t_{123}}$. The delivery of coded subfiles $\{W^{\oplus}_{r_{jk}t_{123}}: j,k\in[3],\ j<k\}$ can be regarded as a \textbf{$3\times 3$ fully cooperative MIMO X-multicast channel}, where every set of two receivers forms a receive multicast group, all the transmitters forms a transmit cooperative group, and the transmit cooperation group has an independent message to send to each receive multicast group.
\begin{lemma}\label{fcX-multicast}
For the $3\times 3$ fully cooperative MIMO X-multicast channel, the achievable DoF per user, denoted as $d_{13}$, is given in \eqref{dofa13}.
\begin{equation}\label{dofa13}
d_{13}=\min \{N, 2M\}
\end{equation}
\end{lemma}

\emph{Proof}: The achievable scheme of this channel is to design the receiver combining matrices. The detailed proof is given in Appendix E.

Based on Lemma~\ref{fcX-multicast}, the NDT of these subfiles is $\frac{2a_{13}}{d_{13}}$.

\subsection{Delivery of Subfiles Cached at Two Receivers and One or More Transmitters}
Consider the delivery of subfiles $\{W_{\kappa r_{k,l}}: \kappa,k,l\in[3],\  k,l\neq \kappa,\ k<l\}$. Since each subfile desired by one receiver is already cached at the other two receivers, we can similarly use coded multicasting. For example, transmitter 1 can generate messages
\begin{align}
 W^{\oplus}_{t_1}  &\triangleq W_{1r_{23}t_{1}}\oplus W_{2r_{13}t_{1}} \oplus W_{3r_{12}t_{1}}, \nonumber \\
 W^{\oplus}_{t_{12}}&   \triangleq W_{1r_{23}t_{12}}\oplus W_{2r_{13}t_{12}} \oplus W_{3r_{12}t_{12}},  \\
 W^{\oplus}_{t_{123}}&  \triangleq W_{1r_{23}t_{123}}\oplus W_{2r_{13}t_{123}} \oplus W_{3r_{12}t_{123}}.\nonumber
\end{align}
Each of the above coded message is desired by all the three receivers, yielding a \textbf{MIMO multicast channel}. We first give the delivery scheme of coded messages $\{ W^{\oplus}_{t_p}: p\in[3]\}$, each with fractional length $a_{21}$. When $N\leq3M$, by \emph{antenna deactivation} \cite{antenna_deactivation}, we let each transmitter use $\frac{N}{3}$ antennas and transmit $\frac{N}{3}$ data streams\footnote{Throughout this paper, if the number of antennas after deactivation or the number of data streams sent from each transmitter (or received by each receiver), denoted as $d$, is not an integer, we can use $t$-symbol extensions such that $td$ is an integer.}. Each user can decode $N$ data streams using $N$ antennas, and the DoF per user is $N$. When $N>3M$, we let each transmitter use $M$ antennas and transmit $M$ data streams. By the antenna deactivation, each user can decode $3M$ data streams using $3M$ antennas, and $3M$ DoF per user can be achieved. For the other two coded messages $\{ W^{\oplus}_{t_{pq}}: p,q\in[3], \ p<q\}$ and $\{ W^{\oplus}_{t_{123}}\}$, we can use the similar scheme as the one used in coded messages $\{ W^{\oplus}_{t_p}: p\in[3]\}$, because each coded message is simultaneously at more than one transmitter. So the DoF per user $d_{21}=d_{22}=d_{23}=\min\{N,3M\}$, and the NDT can be computed as $\tau=\left(\frac{3a_{21}}{d_{21}}+\frac{3a_{22}}{d_{22}}+\frac{a_{23}}{d_{23}}\right)$.

Summing up the NDTs obtained in all the above subsections yields the total NDT  as:
\begin{eqnarray}\label{op}
\begin{split}
\tau=\frac{3a_{01}}{d_{01}}+\frac{3a_{02}}{d_{02}}+\frac{a_{03}}{d_{03}}+\frac{6a_{11}}{d_{11}}+\frac{6a_{12}}{d_{12}}
+\frac{2a_{13}}{d_{13}}+\frac{3a_{21}}{d_{21}}+\frac{3a_{22}}{d_{22}}+\frac{a_{23}}{d_{23}}
\end{split}.
\end{eqnarray}

\section{Optimization of File Splitting Ratios and Connection with Memory Sharing}
 In this section, we study the optimization of the file spitting ratios $\{a_{mn}\}$ to minimize the total NDT in (\ref{op}) subject to the constraints \eqref{1} \eqref{ur} \eqref{ut}. This can be formulated as the following LP problem:
\begin{align}
\mathcal{P}_2:\quad &\min\limits_{\{a_{|{\cal J}||{\cal I}|}\}} \tau(\mu_{R},\mu_{T}) \\
  &\emph{s.t. } \eqref{1} \eqref{ur} \eqref{ut}
\end{align}


Clearly, by defining a new optimization variable $\beta_{mn}$ as:
\begin{equation} \label{29}
\beta_{mn}=\binom{3}{m}\binom{3}{n}a_{mn}, \quad \forall(m,n)\in{\cal A},
\end{equation}
where ${\cal A}=\{(m,n):  m+3n\geq3, m,n\in \{0,1,2,3\}\}$, $\mathcal{P}_2$ can be equivalently expressed as $\mathcal{P}_1$ in Theorem \ref{thm upper}. Here, constraint \eqref{1} is equivalent to constraint \eqref{b1}, and constraints \eqref{ur} and \eqref{ut} are equivalent to constraint \eqref{burut}. By solving $\mathcal{P}_1$, Theorem \ref{thm upper} is then proved.

The significance of rewriting $\mathcal{P}_2$ as $\mathcal{P}_1$ is that $\mathcal{P}_1$ can be interpreted as memory sharing optimization. This is detailed as below.

First, consider an integer point ${\bm \mu}_{mn}^o= \left[\frac{m}{3}, \frac{n}{3}\right]^T$ with $(\mu_R=\frac{m}{3},\mu_T=\frac{n}{3})$ in the cache size region.  Assume that equal file splitting strategy is adopted. That is, each file is split into $\binom{3}{m}\binom{3}{n}a_{mn}$ equal-sized subfiles, each cached simultaneously at $m$ receivers and $n$ transmitters. In that case, we have $a_{mn} = 1/\binom{3}{m}\binom{3}{n}$ and all the rest $a_{m'n'}=0$ . By the delivery scheme introduced in Section IV, the NDT at ${\bm \mu}_{mn}^o$ can be computed as $\tau_{mn}^{o}=\frac{1-m/3}{d_{mn}}$.

Then, consider any feasible point ${\bm \mu}= \left[\mu_R, \mu_T\right]^T$ in the cache size region. The given ${\bm \mu}$ can always be expressed as a convex combination of all the feasible integer points, i.e.,
\begin{equation}\label{convex}
{\bm \mu}=\sum\limits_{(m,n)\in \cal{A}} \beta_{mn}{\bm \mu}_{mn}^{o}.
\end{equation}
We now adopt the memory sharing strategy for cache placement. Namely, we split the transmitter and receiver cache sizes as in \eqref{convex} with memory sharing parameter $\beta_{mn}$. For each $\beta_{mn}$ fraction of the memory, we take $\beta_{mn}$ fraction of each file, split and cache it according to the equal file splitting strategy at the integer point ${\bm \mu}_{mn}^o$. Then, a total achievable NDT can be obtained as
\begin{equation}
\tau=\sum\limits_{(m,n)\in \cal{A}} \beta_{mn}\tau_{mn}^{o}.
\end{equation}
We can minimize the total NDT by finding the optimal memory sharing parameters $\{\beta_{mn}\}$. This is expressed mathematically in $\mathcal{P}_1$ in Theorem \ref{thm upper}.

Both $\mathcal{P}_1$ and $\mathcal{P}_2$ are standard LP problems. By using linear equation substitution and other manipulations, we obtain the closed-form but somewhat tedious expression of the optimal solution $\mu_u$ for any $\mu_R,\mu_T, M, N$ in Appendix A. The antenna configuration is divided into $10$ cases, and for each case the feasible cache size region is partitioned into several regions as shown in Fig. \ref{mul_region}. In each region, the achievable $\tau_{u}$ is a linear decreasing function of $\mu_R$ and $\mu_T$ and hence can be achieved by memory sharing of the integer points within that region.

\textbf{Remark 5:}
In the single antennas case \cite{xu,xu2}, the equal file splitting strategy at integer points is shown to be optimal (in the sense of achieving the optimal solution of the linear programming problem $\mathcal{P}_1$) though not unique. But in the MIMO case, the equal file splitting is not always optimal. For example, consider integer point ${\bm\mu}_{02}^o=[0,\frac{2}{3}]^T$ with $N=5, M=3$, i.e., Case 7 in Fig. \ref{mul_region}.
If equal file splitting strategy is adopted, the NDT is $\frac{1}{2}$. On the other hand, from the optimal solution of $\mathcal{P}_1$ in Theorem \ref{thm upper}, the optimal memory sharing coefficients are $\beta_{01}=\beta_{03}=\frac{1}{2}$. This means that a half of cache size shall be used to adopt the caching scheme at integer point ${\bm\mu}_{01}^o=[0, \frac{1}{3}]^T$ and the other to adopt caching scheme at integer point ${\bm\mu}_{03}^o=[0, 1]^T$. The corresponding NDT shall be  $\frac{1}{2}\cdot\frac{1}{2}+\frac{1}{2}\cdot\frac{1}{3}=\frac{5}{12} < \frac{1}{2}$.

\section{Converse and Multiplicative Gap}
In this section, we present the proof of the NDT lower bound in Theorem \ref{thm lower} and the proof of the maximum multiplicative gap in Corollary \ref{corollary multiplicative gap} .

\subsection{Converse}
We first introduce the following Lemma to help bound the entropy of received signals.
\begin{lemma}\label{entropy_signal}
For the cache-aided MIMO interference network, the differential entropy of the received signals at any $l$ antennas, which can be equipped at different receivers is upper bounded by
\begin{align}
h({\bf y}_{[1:l]})\leq l T\log\Big(2\pi e(cP+1)\Big),
\end{align}
where the parameter $c$ is a function of the channel coefficient.
\end{lemma}

\emph{Proof}: See Appendix F.

Now, we begin the proof. The method of the proof follows the similar cut-set argument in \cite{Alilimits}. Consider any $s\in[3]$ users requesting $\tilde{L}=s\lfloor\frac{L}{s} \rfloor$ different files during $Z=\lfloor\frac{L}{s}\rfloor$ requests. Given the $s$ receivers' caches and received signals during $Z$ requests, these $\tilde{L}$ files can be decoded successfully in the high-SNR regime. Thus, we have:
\begin{subequations}
\begin{align}
F \epsilon_{F} &=H(W_1,\ldots, W_{\tilde{L}} \mid {\bf y}^1_{[1:s]}, \ldots,{\bf y} ^{Z}_{[1:s]}, V_{[1:s]}) \label{converse10}\\
& =H(W_1,\ldots, W_{\tilde{L}})-I(W_1,\ldots, W_{\tilde{L}};{\bf y}^1_{[1:s]}, \ldots,{\bf y} ^{Z}_{[1:s]}, V_{[1:s]}) \\
& =\tilde{L}F-h({\bf y}^1_{[1:s]}, \ldots,{\bf y} ^{Z}_{[1:s]}\mid V_{[1:s]})-H(V_{[1:s]})+h({\bf y}^1_{[1:s]}, \ldots,{\bf y} ^{Z}_{[1:s]}\mid  V_{[1:s]},W_1,\ldots, W_{\tilde{L}}) \nonumber \\
&\quad +H(V_{[1:s]} \mid W_1,\ldots, W_{\tilde{L}}) \\
&\geq \tilde{L}F-h({\bf y}^1_{[1:s]}, \ldots,{\bf y} ^{Z}_{[1:s]})-s\mu_RLF+H(V_{[1:s]} \mid W_1,\ldots, W_{\tilde{L}}) \label{converse11}\\
&\geq \tilde{L}F-h({\bf y}^1_{[1:s]}, \ldots,{\bf y} ^{Z}_{[1:s]})-s\mu_RLF+s(L-\tilde{L})\mu_RF \label{converse12}
\end{align}
\end{subequations}
where \eqref{converse10} follows from the Fano's inequality \cite[Theorem 2.10.1]{cover_it}; \eqref{converse11} comes from the fact that conditioning reduces entropy; \eqref{converse12} comes from the fact that each user caches $\mu_RF$ bits of each file on average and the $s$ receivers know the $\tilde{L}$ files of the total $L$ files.

Using Lemma \ref{entropy_signal}, we further bound  \eqref{converse12} as
\begin{equation}
F \epsilon_{F} \geq \tilde{L}F-sZTN\Theta(\log P)-s\mu_R\tilde{L}F.  \label{converse13}
\end{equation}
%
%
Alternatively, using the data-processing inequality in \cite[Theorem 2.8.1]{cover_it}:
\begin{equation} \label{data_process}
h({\bf x}_{[1:3]}^1, \ldots,{\bf x}_{[1:3]} ^{Z})\geq h({\bf y}^1_{[1:s]}, \ldots,{\bf y} ^{Z}_{[1:s]}),
\end{equation}
we can bound \eqref{converse12} as
\begin{subequations}
\begin{align}
F \epsilon_{F} & \geq \tilde{L}F-h({\bf x}_{[1:3]}^1, \ldots,{\bf x}_{[1:3]} ^{Z})-s\mu_R\tilde{L}F \\
& \geq \tilde{L}F-3ZTM\Theta(\log P)-s\mu_R\tilde{L}F. \label{converse121}
\end{align}
\end{subequations}


Rearranging \eqref{converse13} and \eqref{converse121} and taking $P\rightarrow \infty$ and $F\rightarrow \infty$, the minimum NDT is lower bounded by
\begin{align}\label{lower}
\tau^{\ast} = \lim\limits_{P \to \infty}\lim\limits_{F \to \infty} \frac{T}{F/\log P}& \geq \max\limits_{s\in[3]} \max\left\{\frac{1}{N}(1-s\mu_R),\frac{s}{3M}(1-s\mu_R)\right\} \nonumber \\
&=\max\left\{\frac{1}{N}(1-\mu_R),\max\limits_{s\in[3]}\frac{s}{3M}(1-s\mu_R)\right\},
\end{align}
which completes the proof of Theorem \ref{thm lower}.

\subsection{Multiplicative Gap}
To assist the analysis, we first relax the lower bound $\tau_l$ in \eqref{lower} as follows:
\begin{equation}
\tau_l(\mu_R,\mu_T) \geq \hat {\tau}_l(\mu_R,\mu_T)=
              \begin{cases}
             \frac{1}{N}(1-\mu_R), & \frac{N}{M}\in(0,3]\\
             \frac{1}{3M}(1-\mu_R), &  \frac{N}{M}\in(3,\infty)\\
             \end{cases}.
\end{equation}
The relaxed lower bound can be rewritten as the convex combination at all integer points:
\begin{equation}
\begin{aligned}
\hat{\tau}_l(\mu_R,\mu_T)= & \gamma_{01}\hat{\tau}_l\bigg(0,\frac{1}{3}\bigg)+\gamma_{02}\hat{\tau}_l\left(0,\frac{2}{3}\right)+\gamma_{03}\hat{\tau}_l\bigg(0,1\bigg) +\gamma_{11}\hat{\tau}_l\bigg(\frac{1}{3},\frac{1}{3}\bigg)+\gamma_{12}\hat{\tau}_l\bigg(\frac{1}{3},\frac{2}{3}\bigg) \nonumber \\
& +\gamma_{13}\hat{\tau}_l\bigg(\frac{1}{3},1\bigg)  +\gamma_{21}\hat{\tau}_l\bigg(\frac{2}{3},\frac{1}{3}\bigg) +\gamma_{22}\hat{\tau}_l\bigg(\frac{2}{3},\frac{2}{3}\bigg)+\gamma_{23}\hat{\tau}_l\bigg(\frac{2}{3},1\bigg) \nonumber \\ & +\underbrace{\gamma_{30}\hat{\tau}_l\bigg(1,0\bigg)+\gamma_{31}\hat{\tau}_l\bigg(1,\frac{1}{3}\bigg)+\gamma_{32}\hat{\tau}_l\bigg(1,\frac{2}{3}\bigg)+\gamma_{33}\hat{\tau}_l\bigg(1,1\bigg)}_{=0}
\end{aligned}
\end{equation}
where the combination coefficients satisfy $\sum\limits_{(m,n)\in \cal{A}}\gamma_{mn}=1$ and $\sum\limits_{(m,n)\in \cal{A}}\gamma_{mn}{\bm \mu}_{mn}^{o}=[\mu_R,\mu_T]^T$.

Therefore, we have
\begin{subequations} \label{35}
\begin{align}
\rho \triangleq\frac{\tau_u}{\tau_l} \leq\frac{\tau_u}{\hat\tau_l}&=\frac{\min\limits_{\{\beta_{mn}\}} \left\{ \beta_{01}\tau^o_{01}+\beta_{02}\tau^o_{02}+\cdots+\beta_{23}\tau^o_{23}\right\}}
{\gamma_{01}\hat{\tau}_l(0,1/3)+\gamma_{02}\hat{\tau}_l(0,2/3)+\cdots+\gamma_{23}\hat{\tau}_l(2/3,1)}  \\
&\leq \frac{  \gamma_{01}\tau^o_{01}+\gamma_{02}\tau^o_{02}+\cdots+\gamma_{23}\tau^o_{23}}
{\gamma_{01}\hat{\tau}_l(0,1/3)+\gamma_{02}\hat{\tau}_l(0,2/3)+\cdots+\gamma_{23}\hat{\tau}_l(2/3,1)}   \\
&\leq \max\Big\{\frac{\tau^o_{01}}{\hat{\tau}_l(0,1/3)},\frac{\tau^o_{02}}{\hat{\tau}_l(0,2/3)}, \cdots, \frac{\tau^o_{23}}{\hat{\tau}_l(2/3,1)}    \Big\} \label{sugar}
\end{align}
\end{subequations}
where \eqref{sugar} is due to the inequality $\frac{x_1+x_2+\cdots+x_n}{y_1+y_2+\cdots+y_n}\leq \max\left\{\frac{x_1}{y_1},\frac{x_2}{y_2},\cdots,\frac{x_n}{y_n}\right\}$.

Define $\rho_{mn} \triangleq\frac{\tau^o_{mn}}{\hat{\tau}_l(m/3,n/3)}$. The upper bounds of $\{\rho_{mn}\}$ can be obtained using simple mathematical deduction and are summarized in Table~\ref{table1}. Thus, for any antenna configuration, we have $\rho\leq\max\{\rho_{01},\ldots,\rho_{23}\}\leq 3$, which completes the proof of Corollary \ref{corollary multiplicative gap}.

\begin{table*}
\vspace{-1.0cm}
\centering
\caption{The multiplicative gap at any antenna configurations.}\label{table1}
\begin{tabular}{|c|c|c|c|c|c|c|c|}
\hline
\backslashbox{$\rho$}{$\frac{N}{M}$} & $\left(0,\frac{1}{3}\right]$ & $\left(\frac{1}{3},\frac{2}{3}\right]$ & $\left(\frac{2}{3},1\right]$ & $\left(1,\frac{5}{2}\right]$ & $\left(\frac{5}{2},2\right]$ & $\left(2,3\right]$ & $\left(3,\infty\right)$ \\
\hline
$\rho_{30}\rho_{31}\rho_{32}\rho_{33}$& \multicolumn{7}{c|}{$1$}\\
\hline
$\rho_{21}\rho_{22}\rho_{23}$ &\multicolumn{7}{c|}{$1$}\\
\hline
$\rho_{13}$ &\multicolumn{5}{c|}{$1$} & \multicolumn{2}{c|}{$\frac{3}{2}$} \\
\hline
$\rho_{12}$ &\multicolumn{3}{c|}{$1$} & \multicolumn{4}{c|}{$\frac{3}{2}$} \\
\hline
$\rho_{11}$ & \multicolumn{3}{c|}{$\frac{7}{6}$} & \multicolumn{4}{c|}{$\frac{3}{2}$} \\
\hline
$\rho_{03}$ & \multicolumn{6}{c|}{$1$} & $3$\\
\hline
$\rho_{02}$ & \multicolumn{2}{c|}{$1$} & \multicolumn{3}{c|}{$\frac{5}{2}$} &\multicolumn{2}{c|}{$3$}\\
\hline
$\rho_{01}$ & $1$ & \multicolumn{6}{c|}{$3$}\\
\hline
\end{tabular}
\vspace{-0.7cm}
\end{table*}

\section{Extension to arbitrary number of transmitters and receivers}
 In this section, we discuss the extension of the NDT analysis  to the more general $N_T\times N_R$ networks with $N_T\geq 2$ transmitters and $N_R\geq 2$ receivers.

We first extend the theoretical lower bound on the minimum NDT in the following theorem. The proof is very similar to that of Theorem \ref{thm lower} and hence ignored.
 \begin{theorem}[Lower bound for $N_T\times N_R$ networks]\label{thm general lower}
 Consider the $N_T\times N_R$ cache-aided MIMO interference network where each transmitter is equipped with $M$ antennas and a cache of normalized size $\mu_{T}$, and each receiver is equipped with $N$ antennas and a cache of normalized size $\mu_{R}$. The minimum NDT is lower bounded by
\begin{equation} \label{anylowerbound}
\tau^{\ast} \geq \max\left\{\frac{1}{N}(1-\mu_R),\max\limits_{s\in[N_R]}\frac{s}{N_TM}(1-s\mu_R)\right\}.
\end{equation}
\end{theorem}

Next we discuss the extension of the achievable scheme. By using the same file splitting and caching strategy as in \cite{xu}, an achievable NDT for the $N_T\times N_R$ networks can be similarly expressed in the form of an LP problem as in Theorem \ref{thm upper}:
\begin{subequations}
\begin{align}
\mathcal{P}_3:\quad\tau_{u}\triangleq &  \min \limits_{\{ \beta _{mn}\}} \quad \sum\limits_{(m,n)\in \cal{A}} \beta_{mn}\frac{1-m/N_R}{d_{mn}} \\
  &\ \text{s.t. } \quad \ \sum\limits_{(m,n)\in \cal{A}}  \beta_{mn}=1, \\
  &\quad \quad  \quad \; \sum\limits_{(m,n)\in \cal{A}} \beta_{mn}{\bm \mu}_{mn}^{o}\leq{\bm \mu},\\
  &\quad \quad  \quad \quad 0\leq \beta_{mn}\leq1,\quad \forall (m,n)\in \cal{A}.
\end{align}
\end{subequations}
Here, ${\cal A}=\{(m,n):  m+N_Rn\geq N_R, m\in \{0,1,\cdots,N_R\}, n\in\{0,1,\cdots,N_T\}\}$, and $d_{mn}$ is the DoF per user of the channel formed by the cache state at each integer point ${\bm \mu}^{\circ}_{mn}=\left[\frac{m}{N_R},\frac{n}{N_T}\right]$ in the cache size region. At a general integer point ${\bm \mu}^{\circ}_{mn}$, the newly formed channel is referred to as a $\binom{N_R}{m}\times\binom{N_T}{n}$ cooperative MIMO X-multicast channel, where every set of $m$ out of the total $N_R$ receivers forms a receive multicast group, every set of $n$ out of the total $N_T$ transmitters forms a transmit cooperation group, and each transmit cooperation group has an independent message for each receive multicast group. The DoF per user of this channel can be obtained in some special cases as follows.

$\bullet$ $d_{01}$ at integer point ${\bm \mu}^{\circ}_{01}=\left[0,\frac{1}{N_T}\right]$: In this case, the $\binom{N_R}{m}\times\binom{N_T}{n}$ cooperative MIMO X-multicast channel degenerates to an $N_T\times N_R$ MIMO X channel where each transmitter (receiver) is equipped with $M$ $(N)$ antennas. By using \cite[Theorem 2]{sun_xchannel} and antenna deactivation, an achievable DoF per user of this channel  is given by\footnote{In the general case, we remove the limitation of linear transmission with finite symbol extensions on the achievable schemes.}:
\begin{equation}
d_{01}=\begin{cases}
           \min \left\{\frac{MN_T}{N_R},\frac{qMN_T}{N_T+qN_R-q}\right\}, &{\rm if~}\lfloor\frac{N}{M} \rfloor= q\\
           \min \left\{N,\frac{qNN_T}{qN_T+N_R-q}\right\}, &{\rm if~} \lfloor \frac{M}{N}\rfloor=q
           \end{cases},
\end{equation}
where $q$ is any positive integer.

$\bullet$ $d_{0N_T}$ at integer point ${\bm \mu}^{\circ}_{0N_T}=\left[0,1\right]$: The channel in this case is a MIMO broadcast channel where the virtual transmitter is equipped with $N_TM$ antennas and each receiver is equipped with $N$ antennas. The optimal DoF per user of this channel is
\begin{equation}
d_{0N_T}=
\min\left\{\frac{MN_T}{N_R},N\right\}.
\end{equation}

The DoF results at other integer points remain unknown in general, to our best knowledge. Nevertheless, it is still possible to obtain an achieve NDT based on these special cases. Note that the convex region formed by integer points ${\bm \mu}^{\circ}_{01}$, ${\bm \mu}^{\circ}_{0N_T}$, ${\bm \mu}^{\circ}_{N_R0}$ and ${\bm \mu}^{\circ}_{N_RN_T}$ is the whole cache size region.  Thus, by substituting these four integer points as well as the corresponding DoF values in the LP problem $\mathcal{P}_3$, an achievable NDT for the general $N_T\times N_R$ networks can be obtained as
\begin{equation}\label{upper_any}
\tau^{\ast}\leq\begin{cases}
              \frac{1}{d_{0N_T}}(1-\mu_R), & (\mu_{R},\mu_{T})\in {\cal R}_{1} \\
              \left[\frac{1}{d_{0N_T}}-\frac{N_T}{N_T-1}\left(\frac{1}{d_{0N_T}}-\frac{1}{d_{01}}\right) \right](1-\mu_R)-\frac{N_T}{N_T-1}\left(\frac{1}{d_{01}}-\frac{1}{d_{0N_T}}\right)\mu_T,                                & (\mu_{R},\mu_{T})\in {\cal R}_{2}
              \end{cases},
\end{equation}
where
\begin{equation}
\begin{cases}
 {\cal R}_{1}=\{(\mu_{R},\mu_{T}): \mu_{R}+\mu_{T}\geq 1,\mu_{R}\leq 1,\mu_{T}\leq 1\} \\
 {\cal R}_{2}=\{(\mu_{R},\mu_{T}): \mu_{R}+\mu_{T}< 1,\mu_{R}\geq0, \mu_{R}+N_T\mu_{T}\geq 1\}
 \end{cases}.
 \end{equation}

By comparing \eqref{upper_any} with \eqref{anylowerbound}, it is found that the achievable NDT is optimal when (1) $\frac{N}{M}\in\left(0,\frac{1}{N_R}\right]$, and (2) $\frac{N}{M}\in\left(\frac{1}{N_R},\frac{N_T}{N_R}\right]$ and $(\mu_R,\mu_T)\in {\cal R}_{1}$. It is also found that the multiplicative gap $\rho$ is less than $2$ when $N_T\geq N_R$ and $N=M$. In a more general setting of $(M, N, N_T, N_R)$, the gap does not converge to a constant.  Further investigation is needed but beyond the scope of this paper.

\section{conclusion}
In this paper, we study the storage-latency tradeoff in the $3\times 3$ cache-aided MIMO interference network. With different file splitting patterns, the MIMO interference channel can be turned to MIMO broadcast channel, MIMO multicast channel, MIMO X channel, or hybrid forms of these channels. We propose linear transmission schemes and obtain the DoF results of these channels. We obtain the achievable upper bound of minimum NDT by solving a linear programming problem. The achievable NDT decreases piecewise linearly with the normalized cache sizes and each additive item is inversely proportional to the number of antennas. This finding  reveals that the MIMO gain and cache gain are cumulative in the considered wireless network. We also give a lower bound of minimum NDT. It is shown that the achievable NDT is optimal in certain cases and is within a multiplicative gap of 3 to the optimal in other cases.

Although this work mainly focuses on the network with three transmitters and three receivers, the results have been extended, in certain ways, to the more general network with arbitrary number of transmitters and receivers.  The main challenge for further investigation would be the DoF analysis of the new class of cooperative MIMO X-multicast channels.


\begin{small}

\section*{Appendix A: the closed form expression of $\tau_{u}$ in Theorem \ref{thm upper}}
\emph{Case 1}: $\frac{N}{M}\in \left(0, \frac{1}{3}\right]$
\begin{equation}
\tau_u(\mu_{R},\mu_{T})= \frac{1}{N}\left(1-\mu_{R}\right)
\end{equation}

\emph{Case 2}: $\frac{N}{M}\in \left(\frac{1}{3}, \frac{4}{9}\right]$
\begin{equation}
\tau_u(\mu_{R},\mu_{T})=
                                 \begin{cases}
                                   \frac{1}{N}\left(1-\mu_{R}\right) & (\mu_{R},\mu_{T})\in {\cal R}_{1}\\
                                  \frac{1}{N}\left(-1+\mu_{R}+3\mu_{T}\right)+\frac{3}{M}\left(2-2\mu_{R}-3\mu_{T}\right) & (\mu_{R},\mu_{T})\in {\cal R}_{2} \\
                                 \end{cases}
\end{equation}

\emph{Case 3}: $\frac{N}{M}\in \left(\frac{4}{9}, \frac{2}{3}\right]$
\begin{equation}
\tau_u(\mu_{R},\mu_{T})=
                                 \begin{cases}
                                   \frac{1}{N}\left(1-\mu_{R}\right) & (\mu_{R},\mu_{T})\in {\cal R}_{1}\\
                                   \frac{1}{3N}\left(5-5\mu_{R}-3\mu_{T}\right) & (\mu_{R},\mu_{T})\in {\cal R}_{2} \\
                                  \frac{1}{\min\{\frac{N}{2-1/ \xi},\frac{M}{2+1/\xi}\}}\left(2-3\mu_{R}-3\mu_{T}\right)+\frac{1}{3N}\left(-3+7\mu_{R}+9\mu_{T}\right) & (\mu_{R},\mu_{T})\in {\cal R}_{3} \\
                                  \frac{3}{\min\{\frac{N}{2-1/ \xi},\frac{M}{2+1/\xi}\}}\left(1-\mu_{R}-2\mu_{T}\right)+\frac{7}{3N}\left(-1+\mu_{R}+3\mu_{T}\right) & (\mu_{R},\mu_{T})\in {\cal R}_{4} \\
                                 \end{cases}
\end{equation}

\emph{Case 4}: $\frac{N}{M}\in \left(\frac{2}{3}, \frac{20}{27}\right]$
\begin{equation}
\tau_u(\mu_{R},\mu_{T})=
                                 \begin{cases}
                                   \frac{1}{N}\left(1-\mu_{R}\right) & (\mu_{R},\mu_{T})\in {\cal R}_{1}\\
                                   \frac{3}{N}(1-\mu_R-\mu_T) +\frac{3}{2M}\left(-2+2\mu_{R}+3\mu_T\right)& (\mu_{R},\mu_{T})\in {\cal R}_{2} \\
                                  \frac{7}{3N}\left(2-2\mu_{R}-3\mu_{T}\right)+\frac{9}{2M}\left(-1+\mu_{R}+2\mu_{T}\right) & (\mu_{R},\mu_{T})\in {\cal R}_{3} \\
                                  \frac{1}{\min\{\frac{N}{2-1/ \xi},\frac{M}{2+1/\xi}\}}\left(2-3\mu_{R}-3\mu_{T}\right)+\frac{7}{3N}\mu_{R}+\frac{3}{2M}\left(-1+3\mu_{T}\right) & (\mu_{R},\mu_{T})\in {\cal R}_{4} \\
                                  \frac{3}{\min\{\frac{N}{2-1/ \xi},\frac{M}{2+1/\xi}\}}\left(1-\mu_{R}-2\mu_{T}\right)+\frac{7}{3N}\left(-1+\mu_{R}+3\mu_{T}\right) & (\mu_{R},\mu_{T})\in {\cal R}_{5} \\
                                 \end{cases}
\end{equation}

\emph{Case 5}: $\frac{N}{M}\in \left(\frac{20}{27}, 1\right]$
\begin{equation}\label{48}
\tau_u(\mu_{R},\mu_{T})=
                                 \begin{cases}
                                   \frac{1}{N}\left(1-\mu_{R}\right) & (\mu_{R},\mu_{T})\in {\cal R}_{1}\\
                                   \frac{1}{3N}\left(4-4\mu_R-\mu_{T}\right)& (\mu_{R},\mu_{T})\in {\cal R}_{2} \\
                                  \frac{7}{3N}\mu_R+\frac{9}{2M}\left(1-2\mu_{R}-\mu_{T}\right)+\frac{1}{N}\left(-2+3\mu_{R}+3\mu_{T}\right) & (\mu_{R},\mu_{T})\in {\cal R}_{3} \\
                                  \frac{1}{\min\{\frac{N}{2-1/ \xi},\frac{M}{2+1/\xi}\}}\left(2-3\mu_{R}-3\mu_{T}\right)+\frac{7}{3N}\mu_{R}+\frac{3}{2M}\left(-1+3\mu_{T}\right) & (\mu_{R},\mu_{T})\in {\cal R}_{4} \\
                                   \frac{3}{\min\{\frac{N}{2-1/ \xi},\frac{M}{2+1/\xi}\}}\left(1-\mu_{R}-2\mu_{T}\right)+\frac{7}{3N}\left(-1+\mu_{R}+3\mu_{T}\right) & (\mu_{R},\mu_{T})\in {\cal R}_{5} \\
                                 \end{cases}
\end{equation}

\emph{Case 6}: $\frac{N}{M}\in \left(1, \frac{4}{3}\right]$
\begin{equation}
\tau_u(\mu_{R},\mu_{T})=
                                 \begin{cases}
                                   \frac{1}{N}\left(1-\mu_{R}\right) & (\mu_{R},\mu_{T})\in {\cal R}_{1}\\
                                   \frac{1}{2N}\left(-3+4\mu_R+3\mu_T\right)+\frac{1}{2M}(5-6\mu_R-3\mu_T)& (\mu_{R},\mu_{T})\in {\cal R}_{2} \\
                                  \frac{2}{\max\{\frac{6}{7}M,\frac{2}{3}N\}}\left(1-\mu_{R}-\mu_{T}\right)+\frac{1}{2N}\left(-1+2\mu_{R}+\mu_{T}\right) +\frac{1}{2M}\left(-1+3\mu_{T}\right) & (\mu_{R},\mu_{T})\in {\cal R}_{3} \\
                                  \frac{2}{\max\{\frac{6}{7}M,\frac{2}{3}N\}}\mu_R+\frac{3}{2\min\{\frac{M}{2-1/ \xi},\frac{N}{2+1/\xi}\}}\left(1-2\mu_{R}-\mu_{T}\right)+\frac{1}{2M}\left(-1+3\mu_{T}\right) & (\mu_{R},\mu_{T})\in {\cal R}_{4} \\
                                  \frac{2}{\max\{\frac{6}{7}M,\frac{2}{3}N\}}\left(-1+\mu_{R}+3\mu_{T}\right)+\frac{3}{\min\{\frac{M}{2-1/ \xi},\frac{N}{2+1/\xi}\}}\left(1-\mu_{R}-2\mu_{T}\right) & (\mu_{R},\mu_{T})\in {\cal R}_{5} \\
                                   \frac{2}{\max\{\frac{6}{7}M,\frac{2}{3}N\}}\left(1-\mu_{R}-\mu_{T}\right)+\frac{1}{N}(-1+\mu_R+2\mu_T) & (\mu_{R},\mu_{T})\in {\cal R}_{6}\\
                                 \end{cases}
\end{equation}

\emph{Case 7}: $\frac{N}{M}\in \left(\frac{4}{3}, 2\right]$
\begin{equation}
\tau_u(\mu_{R},\mu_{T})=
                                 \begin{cases}
                                   \frac{1}{N}\left(1-\mu_{R}\right) & (\mu_{R},\mu_{T})\in {\cal R}_{1}\\
                                   \frac{1}{6N}\left(11-12\mu_R-3\mu_T\right)& (\mu_{R},\mu_{T})\in {\cal R}_{2} \\
                                  \frac{1}{2N}\left(1+4\mu_R-\mu_T\right)+\frac{1}{M}\left(1-3\mu_R\right) & (\mu_{R},\mu_{T})\in {\cal R}_{3} \\
                                  \frac{3}{N}\mu_R+\frac{1}{4M}\left(7-18\mu_R-3\mu_T\right)& (\mu_{R},\mu_{T})\in {\cal R}_{4} \\
                                  \frac{3}{N}\left(-1+\mu_R+3\mu_T\right)+\frac{9}{2M}\left(1-\mu_R-2\mu_T\right) & (\mu_{R},\mu_{T})\in {\cal R}_{5} \\
                                  \frac{1}{N}(2-2\mu_R-\mu_T) & (\mu_{R},\mu_{T})\in {\cal R}_{6}\\
                                 \end{cases}
\end{equation}

\emph{Case 8}: $\frac{N}{M}\in \left(2, \frac{12}{5}\right]$
\begin{equation}
\tau_u(\mu_{R},\mu_{T})=
                                 \begin{cases}
                                   \frac{1}{N}\left(1-\mu_{R}\right) & (\mu_{R},\mu_{T})\in {\cal R}_{1}\\
                                   \frac{1}{3N}\left(-1+3\mu_R\right)+\frac{1}{3M}\left(2-3\mu_R\right)& (\mu_{R},\mu_{T})\in {\cal R}_{2} \\
                                  \frac{1}{6N}\left(13-12\mu_R-9\mu_T\right)+\frac{1}{6M}\left(-1+3\mu_T\right) & (\mu_{R},\mu_{T})\in {\cal R}_{3} \\
                                  \frac{3}{2N}\left(1-\mu_T\right)+\frac{1}{2M}\left(1-4\mu_T+\mu_T\right) & (\mu_{R},\mu_{T})\in {\cal R}_{4} \\
                                  \frac{3}{2N}\left(5-8\mu_R-5\mu_T\right)+\frac{1}{M}\left(-2+3\mu_R+3\mu_T\right)& (\mu_{R},\mu_{T})\in {\cal R}_{5} \\
                                  \frac{1}{2N}\left(7-12\mu_R-3\mu_T\right) & (\mu_{R},\mu_{T})\in {\cal R}_{6} \\
                                  \frac{3}{N}(2-2\mu_R-3\mu_T) & (\mu_{R},\mu_{T})\in {\cal R}_{7}\\
                                  \frac{1}{N}(2-2\mu_R-\mu_T) & (\mu_{R},\mu_{T})\in {\cal R}_{8}\\
                                 \end{cases}
\end{equation}

\emph{Case 9}: $\frac{N}{M}\in \left(\frac{12}{5}, 3\right]$
\begin{equation}
\tau_u(\mu_{R},\mu_{T})=
                                 \begin{cases}
                                   \frac{1}{N}\left(1-\mu_{R}\right) & (\mu_{R},\mu_{T})\in {\cal R}_{1}\\
                                   \frac{1}{3N}\left(-1+3\mu_R\right)+\frac{1}{3M}\left(2-3\mu_R\right)& (\mu_{R},\mu_{T})\in {\cal R}_{2} \\
                                   \frac{1}{6N}\left(13-12\mu_R-9\mu_T\right)+\frac{1}{6M}\left(-1+3\mu_T\right)& (\mu_{R},\mu_{T})\in {\cal R}_{3} \\
                                  \frac{1}{\min\{\frac{2}{5}N,M\}}\left(1-3\mu_R\right)+\frac{1}{2N}\left(2+3\mu_R-3\mu_T\right)+\frac{1}{6M}\left(-2+3\mu_R+3\mu_T\right) & (\mu_{R},\mu_{T})\in {\cal R}_{4}\\
                                  \frac{3}{\min\{\frac{2}{5}N,M\}}\left(1-\mu_T\right)+\frac{1}{M}\left(-2-2\mu_R+3\mu_T\right) & (\mu_{R},\mu_{T})\in {\cal R}_{5}\\
                                  \frac{3}{N}\left(2-2\mu_R-3\mu_T\right)+\frac{1}{M}\left(-1+3\mu_T\right)& (\mu_{R},\mu_{T})\in {\cal R}_{6} \\
                                  \frac{3}{N}(2-2\mu_R-3\mu_T) & (\mu_{R},\mu_{T})\in {\cal R}_{7}\\
                                  \frac{1}{N}(2-2\mu_R-\mu_T) & (\mu_{R},\mu_{T})\in {\cal R}_{8}\\
                                 \end{cases}
\end{equation}

\emph{Case 10}: $\frac{N}{M}\in \left(3, \infty\right]$
\begin{equation}
\tau_u(\mu_{R},\mu_{T})=
                                 \begin{cases}
                                  \frac{1}{3M}\left(1-\mu_R\right) & (\mu_{R},\mu_T)\in {\cal R}_{1}\\
                                   \frac{1}{9M}\left(5-6\mu_R\right)& (\mu_{R},\mu_T)\in {\cal R}_{2} \\
                                   \frac{1}{M}\left(1-2\mu_R\right) & (\mu_{R},\mu_T)\in {\cal R}_{3} \\
                                  \frac{1}{M}\left(2-2\mu_R-3\mu_T\right)& (\mu_R,\mu_T)\in {\cal R}_{4} \\
                                  \frac{1}{3M}(2-2\mu_R-\mu_T) & (\mu_R,\mu_T)\in {\cal R}_{5}\\
                                 \end{cases}
\end{equation}
\begin{figure*}[t]
 \vspace{-0.7cm}
\begin{centering}
\includegraphics[scale=0.8]{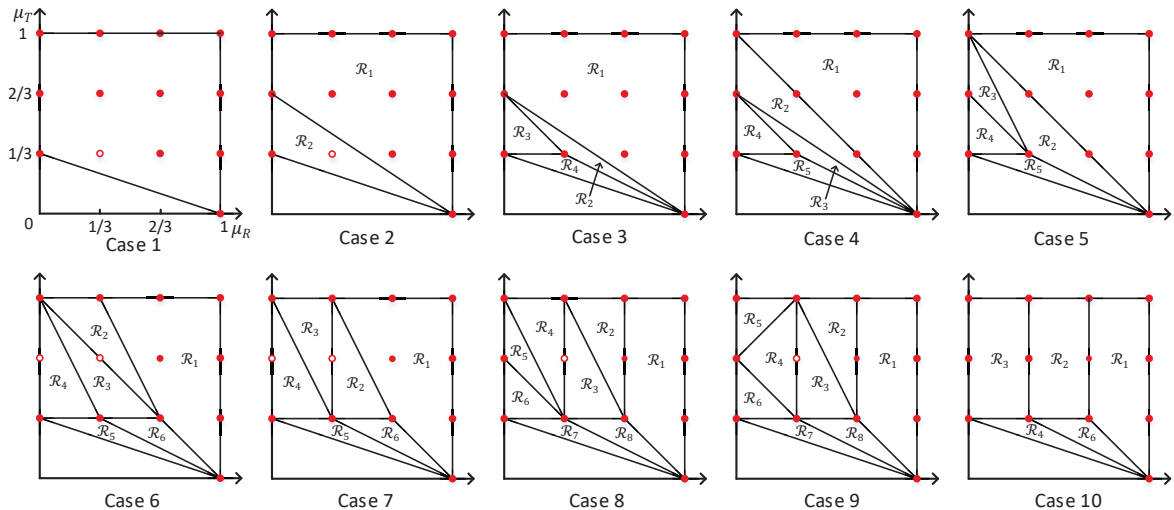}
 \caption{Cache regions of the different number of antennas.}\label{mul_region}
\end{centering}
\end{figure*}
The cache size regions $\{{\cal R}_i \}$ of each case are illustrated in Fig. \ref{mul_region}.

\section*{Appendix B: proof of Lemma~\ref{cooperativeX}}
Throughout this Appendix and Appendices C, D and E, we adopt the DoF plane introduced in \cite{Liu4} to present the DoF results.
The DoF per user of the $3\times 3$ partially cooperative MIMO X channel shown in \eqref{dofa02} of Lemma~\ref{cooperativeX}  is illustrated in Fig. \ref{dtotal}(a). To prove its achievability, it suffices to prove the achievability of points $\{Q_1, Q_2\}$ in the DoF plane, by \cite[Lemma 2]{Liu4}.

The achievable scheme of the $3\times 3$ partially cooperative MIMO X channel needs three phases as shown in Fig. \ref{a02}. In each phase, the transmission scheme is similar and we take the phase I for an example. Let the $d\times 1$ vectors ${\bf s}_{r_1t_{12}}$, ${\bf s}_{r_2t_{13}}$, ${\bf s}_{r_3t_{23}}$ denote the actual transmitted signal vectors of messages $A_{12}$, $B_{13}$, $C_{23}$, intended for receivers 1, 2, and 3, respectively. Here, $d$ is the desired DoF per user. Due to the symmetry of the three receivers, we take receiver 1 as an example. Its received signal (ignoring noise for brevity) can be written as
\begin{equation}
\begin{aligned}
  {\bf y}_1=&{\bf H}_{11}({\bf V}_{r_{1}t_{12}1}{\bf s}_{r_1t_{12}}+{\bf V}_{r_{2}t_{13}1}{\bf s}_{r_2t_{13}})+{\bf H}_{12}({\bf V}_{r_{1}t_{12}2}{\bf s}_{r_1t_{12}}+{\bf V}_{r_{3}t_{23}2}{\bf s}_{r_3t_{23}})\\
  &+{\bf H}_{13}({\bf V}_{r_{2}t_{13}3}{\bf s}_{r_2t_{13}}+{\bf V}_{r_{3}t_{23}3}{\bf s}_{r_3t_{23}}),
\end{aligned}
\end{equation}
where ${\bf V}_{r_{k}t_{pq}i}$ is the $M\times d$ precoding matrix of signal ${\bf s}_{r_{k}t_{pq}}$ at transmitter $i\in\{p,q\}$. Next, we give the detailed design method of transmitter precoding matrices and, if necessarily, receiver combining matrices.

\emph{(1) the achievability of $Q_1$}: This is to show that the DoF per user $d=\frac{2M}{3}$ is achievable at antenna configuration $N=\frac{2M}{3}$. For receiver $1$, ${\bf s}_{r_{2}t_{13}}$ and ${\bf s}_{r_{3}t_{23}}$ are the interference signals. We can design the $M\times \frac{2M}{3}$ precoding matrices ${\bf V}_{r_{2}t_{13}1}$, ${\bf V}_{r_{2}t_{13}3}$, ${\bf V}_{r_{3}t_{23}2}$ and ${\bf V}_{r_{3}t_{23}3}$ to satisfy:
\begin{equation}\label{P1}
              \begin{array}{c}
                 {\bf H}_{11}{\bf V}_{r_{2}t_{13}1}=-{\bf H}_{13}{\bf V}_{r_{2}t_{13}3} \\
                 {\bf H}_{12}{\bf V}_{r_{3}t_{23}2}=-{\bf H}_{13}{\bf V}_{r_{3}t_{23}3}
                \end{array}
\end{equation}
In this way, the interferences from ${\bf s}_{r_{2}t_{13}}$ and ${\bf s}_{r_{3}t_{23}}$ will be cancelled at receiver 1, which is known as interference neutralization. Similarly, the interferences at receiver 2 and 3 can be neutralized by the following design:
\begin{eqnarray}\label{P2P3}
\begin{split}
 & {\bf H}_{21}{\bf V}_{r_{1}t_{12}1}=-{\bf H}_{22}{\bf V}_{r_{1}t_{12}2},\quad {\bf H}_{22}{\bf V}_{r_{3}t_{23}2}=-{\bf H}_{23}{\bf V}_{r_{3}t_{23}3}, \\
 & {\bf H}_{31}{\bf V}_{r_{2}t_{13}1}=-{\bf H}_{33}{\bf V}_{r_{2}t_{13}3},\quad {\bf H}_{31}{\bf V}_{r_{1}t_{12}1}=-{\bf H}_{32}{\bf V}_{r_{1}t_{12}2}.
\end{split}
\end{eqnarray}
Note that each precoding matrix needs to meet two conditions from \eqref{P1} and \eqref{P2P3}. The existence is justified as follows. We take ${\bf V}_{r_{1}t_{12}1}$ and ${\bf V}_{r_{1}t_{12}2}$ as examples. They can be designed as:
\begin{equation}\label{va1va2}
\left[\begin{array}{c}
           {\bf V}_{r_{1}t_{12}1} \\
           {\bf V}_{r_{1}t_{12}2} \\
         \end{array}
       \right]\subseteq \text{null}\left[\begin{array}{c}
    {\bf H}_{21} \quad   {\bf H}_{22} \\
    {\bf H}_{31} \quad   {\bf H}_{32} \\
  \end{array}
\right].
\end{equation}
Given that each ${\bf H}_{ji}$ is a $\frac{2M}{3}\times M$ matrix, using the null space theorem, we can obtain such precoding matrices ${\bf V}_{r_{1}t_{12}1}$ and ${\bf V}_{r_{1}t_{12}2}$ that satisfy \eqref{P2P3} with probability one. In this way, all the interferences are cancelled and each receiver can decode $\frac{2M}{3}$ data streams.

\emph{(2) the achievability of $Q_2$}:
 We intend to achieve DoF per user $d=M$ at antenna configuration $N=\frac{5M}{2}$. In this case, we need to jointly design the transmit precoding matrices and the receive combining matrices for interference neutralization. In specific, we first design the $M\times \frac{5}{2}M$ combining matrices, denoted as ${\bf P}_{j}$, for each receiver $j$ as follows:
\begin{equation}
\begin{aligned}
\left[\begin{array}{c}
       {\bf p}_1^1 \\
       {\bf p}_1^2 \\
       \vdots \\
       {\bf p}_1^\frac{M}{2}
     \end{array}\right]^{T}\subseteq\text{null}
     \left[  \begin{array}{c}
               {\bf H}_{11}^T \\
               {\bf H}_{13}^T
             \end{array}
     \right], \quad \left[\begin{array}{c}
       {\bf p}_1^{\frac{M}{2}+1} \\
       {\bf p}_1^{\frac{M}{2}+2} \\
       \vdots \\
       {\bf p}_1^{M}
     \end{array}\right]^{T}\subseteq\text{null}
     \left[  \begin{array}{c}
               {\bf H}_{12}^T \\
               {\bf H}_{13}^T
             \end{array}
     \right],\quad
     \left[\begin{array}{c}
       {\bf p}_2^1 \\
       {\bf p}_2^2 \\
       \vdots \\
       {\bf p}_2^\frac{M}{2}
     \end{array}\right]^{T}\subseteq\text{null}
     \left[  \begin{array}{c}
               {\bf H}_{21}^T \\
               {\bf H}_{22}^T
             \end{array}
     \right], \\
      \left[\begin{array}{c}
       {\bf p}_2^{\frac{M}{2}+1} \\
       {\bf p}_2^{\frac{M}{2}+2} \\
       \vdots \\
       {\bf p}_2^{M}
     \end{array}\right]^{T}\subseteq\text{null}
     \left[  \begin{array}{c}
               {\bf H}_{22}^T \\
               {\bf H}_{23}^T
             \end{array}
     \right],\quad
     \left[\begin{array}{c}
       {\bf p}_3^1 \\
       {\bf p}_3^2 \\
       \vdots \\
       {\bf p}_3^\frac{M}{2}
     \end{array}\right]^{T}\subseteq\text{null}
     \left[  \begin{array}{c}
               {\bf H}_{31}^T \\
               {\bf H}_{32}^T
             \end{array}
     \right], \quad \left[\begin{array}{c}
       {\bf p}_3^{\frac{M}{2}+1} \\
       {\bf p}_3^{\frac{M}{2}+2} \\
       \vdots \\
       {\bf p}_3^{M}
     \end{array}\right]^{T}\subseteq\text{null}
     \left[  \begin{array}{c}
               {\bf H}_{31}^T \\
               {\bf H}_{33}^T
             \end{array}
     \right],
\end{aligned}
\end{equation}
where ${\bf p}_j^m$ denotes the $m$-th row of ${\bf P}_j$.

Then, we design the $M\times M$ transmit precoding matrices to meet the same conditions in (\ref{P1}) and (\ref{P2P3}) with each channel matrix ${\bf H}_{ji}$ replaced by the effective channel matrix ${\bf P}_j{\bf H}_{ji}$. The existence of such precoding matrices is justified using the similar null space theorem as in \eqref{va1va2}.
%
%
By doing so, all the interferences are neutralized and $M$ data streams can be decoded by each receiver.

\section*{Appendix C: proof of Lemma~\ref{X-multicast}}
The DoF per user of the $3\times 3$ MIMO X-multicast channel, shown in \eqref{dofa11} of Lemma~\ref{X-multicast}, is illustrated in Fig. \ref{dtotal}(b).
 \begin{figure*}[t]
\begin{centering}
\includegraphics[scale=0.5]{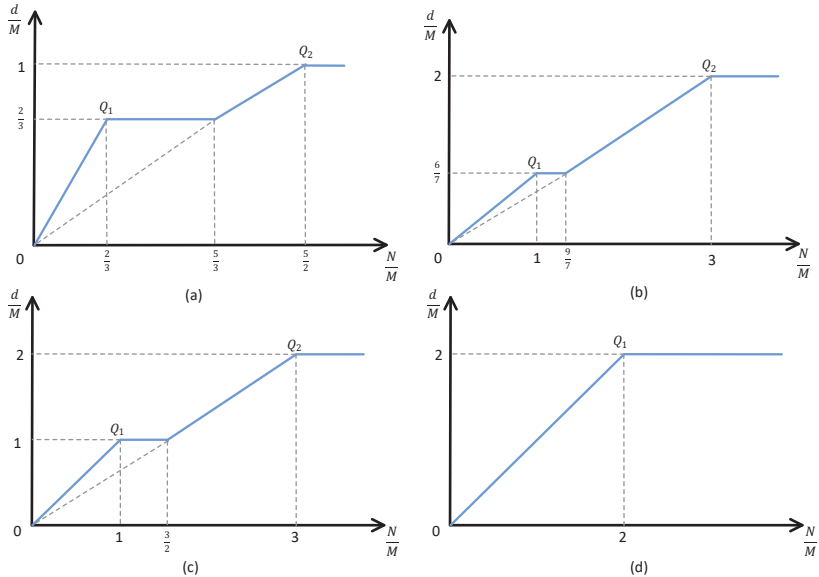}
 \caption{The DoF planes: (a) the $3\times 3$ partially cooperative MIMO X channel, (b) the $3\times 3$ MIMO X-multicast channel, (c) the $3\times 3$ partially cooperative MIMO X-multicast channel, (d) the $3\times 3$ fully cooperative MIMO X-multicast channel. }\label{dtotal}
\end{centering}
\end{figure*}

%

\emph{(1) the achievability of $Q_1$}:
 This is to show that the DoF per user $d=\frac{6M}{7}$ is achievable at antenna configuration $N=M$.
 Let ${\bf s}_{r_{jk}t_{p}}$ denote the $M\times \frac{M}{7}$ transmitted signal vector for $ W^{\oplus}_{r_{jk}t_p}$, intended to receive multicast group $\{j,k\}$ from transmitter $p$. Let ${\bf V}_{r_{jk}t_{p}}$ denote the $M\times \frac{M}{7}$ precoding matrix of ${\bf s}_{r_{jk}t_{p}}$ at transmitter $p$. At receiver $1$, the received signal can be expressed as (ignoring the noise for brevity)
 \begin{align}
 {\bf y}_{1}={\bf H}_{11}({\bf V}_{r_{12}t_{1}}{\bf s}_{r_{12}t_{1}}+{\bf V}_{r_{23}t_{1}}{\bf s}_{r_{23}t_{1}}+{\bf V}_{r_{13}t_{1}}{\bf s}_{r_{13}t_{1}}) \nonumber\\
 +{\bf H}_{12}({\bf V}_{r_{12}t_{2}}{\bf s}_{r_{12}t_{2}}+{\bf V}_{r_{23}t_{2}}{\bf s}_{r_{23}t_{2}}+{\bf V}_{r_{13}t_{2}}{\bf s}_{r_{13}t_{2}})\\
 +{\bf H}_{13}({\bf V}_{r_{12}t_{3}}{\bf s}_{r_{12}t_{3}}+{\bf V}_{r_{23}t_{3}}{\bf s}_{r_{23}t_{3}}+{\bf V}_{r_{13}t_{3}}{\bf s}_{r_{13}t_{3}}). \nonumber
 \end{align}
 Receiver 1 desires signals ${\bf s}_{r_{12}t_{1}},{\bf s}_{r_{13}t_{1}},{\bf s}_{r_{12}t_{2}},{\bf s}_{r_{13}t_{2}},{\bf s}_{r_{12}t_{3}}$, and ${\bf s}_{r_{13}t_{3}}$, and it wants to align the interference signals ${\bf s}_{r_{23}t{_1}},{\bf s}_{r_{23}t{_2}}$, and ${\bf s}_{r_{23}t{_3}}$ along a same direction so as to cancel them all at once:
 \begin{align}
 {\bf H}_{11}{\bf V}_{r_{23}t_{1}}={\bf H}_{12}{\bf V}_{r_{23}t_{2}}={\bf H}_{13}{\bf V}_{r_{23}t_{3}}\triangleq{\bf V}_{1}.
 \end{align}
At receiver 2 and 3, the similar equations can be obtained:
 \begin{align}
 {\bf H}_{21}{\bf V}_{r_{13}t_{1}}={\bf H}_{22}{\bf V}_{r_{13}t_{2}}={\bf H}_{23}{\bf V}_{r_{13}t_{3}}\triangleq{\bf V}_{2}, \\
 {\bf H}_{31}{\bf V}_{r_{12}t_{1}}={\bf H}_{32}{\bf V}_{r_{12}t_{2}}={\bf H}_{33}{\bf V}_{r_{12}t_{3}}\triangleq{\bf V}_{3}.
 \end{align}
We need to further design ${\bf V}_{1}$, ${\bf V}_{2}$ and ${\bf V}_{3}$ to ensure the decodability of desired signals at each receiver. We give an achievable method as below:
\begin{equation}
{\bf V}_{1}={\bf V}_{2}={\bf V}_{3}=\underbrace{\text{diag}\{{\bf 1}_{7 \times 1},{\bf 1}_{7 \times 1}, \cdots ,{\bf 1}_{7 \times 1}\}}_{M\times \frac{M}{7}},
\end{equation}
where ${\bf 1}_{7 \times 1}$ denotes the $7 \times 1$ vector with all elements being one. In this way, all desired signals are linearly independent of each other and each receiver can decode its desired signals successfully. So the $\frac{6M}{7}$ DoF per user can be obtained.

\begin{figure*}[t]
\begin{centering}
\includegraphics[scale=0.7]{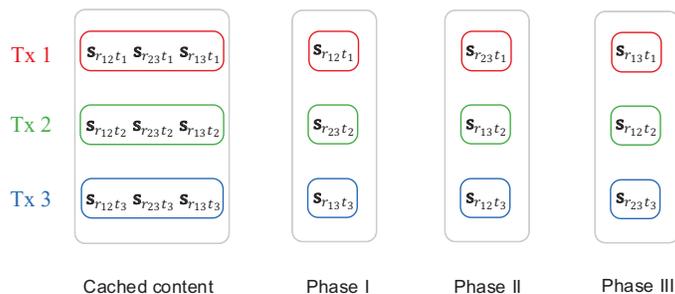}
 \caption{The alternating transmission scheme in the $3\times 3$ MIMO  X-multicast channel.}\label{ala11}
\end{centering}
\end{figure*}
\emph{(2) the achievability of $Q_2$}: We intend to achieve DoF per user $d=2M$ at antenna configuration $N=3M$. In this case, we use the alternating transmission scheme as shown in Fig. \ref{ala11}.
We take phase I as an example. For receiver 1, the post-processed received signal (ignoring the noise for brevity) after the $2M \times 3M$ combining matrix ${\bf P}_1$ can be expressed as
 \begin{align}
 {\bf \hat{y}}_{1}={\bf P}_1({\bf H}_{11}{\bf s}_{r_{12}t_{1}}+{\bf H}_{12}{\bf s}_{r_{23}t_{2}}+{\bf H}_{13}{\bf s}_{r_{13}t_{3}}).
 \end{align}
 We can design the combining matrix ${\bf P}_1$ to cancel the interference signal ${\bf s}_{r_{23}t_{2}}$ as
 \begin{equation}
 {\bf P}_1^T\subseteq\text{null}\left[{\bf H}_{12}^{T}\right].
 \end{equation}
 The other combining matrices can be designed in the similar way. Thus, each receiver can obtain its desired signals without interference and hence $2M$ DoF per user is achievable.

\section*{Appendix D: proof of Lemma~\ref{pcoperativeX-multicast}}
The DoF per user of the $3\times 3$ partially cooperative MIMO X-multicast channel, shown in \eqref{dofa12} of Lemma~\ref{pcoperativeX-multicast}, is illustrated in Fig. \ref{dtotal}(c).

\emph{(1) the achievability of $Q_1$}: This is to show that the DoF per user $d=M$ is achievable at antenna configuration $N=M$.
 Let ${\bf s}_{r_{jk}t_{pq}}$ be the $\frac{M}{6}\times 1$ transmitted signal vector for $W^{\oplus}_{r_{jk}t_{pq}}$, intended to receive multicast group $\{j,k\}$ from transmitter cooperation group $\{p,q\}$. Let ${\bf V}_{r_{jk}t_{pq}i}$ be the $M\times \frac{M}{6}$ precoding matrix of ${\bf s}_{r_{jk}t_{pq}}$ at transmitter $i\in\{p,q\}$. The received signal (ignoring the noise for brevity) at receiver $1$ can be expressed as
 \begin{eqnarray}
 \begin{split}
 {\bf y}_{1}
 &=({\bf H}_{11}{\bf V}_{r_{12}t_{12}1}+{\bf H}_{12}{\bf V}_{r_{12}t_{12}2}){\bf s}_{r_{12}t_{12}}+({\bf H}_{11}{\bf V}_{r_{23}t_{12}1}+{\bf H}_{12}{\bf V}_{r_{23}t_{12}2}){\bf s}_{r_{23}t_{12}}  \nonumber \\
 &+({\bf H}_{11}{\bf V}_{r_{13}t_{12}1}+{\bf H}_{12}{\bf V}_{r_{13}t_{12}2}){\bf s}_{r_{13}t_{12}}+({\bf H}_{11}{\bf V}_{r_{12}t_{13}1}+{\bf H}_{13}{\bf V}_{r_{12}t_{13}3}){\bf s}_{r_{12}t_{13}}  \nonumber \\
 &+({\bf H}_{11}{\bf V}_{r_{23}t_{13}1}+{\bf H}_{13}{\bf V}_{r_{23}t_{13}3}){\bf s}_{r_{23}t_{13}}+({\bf H}_{11}{\bf V}_{r_{13}t_{13}1}+{\bf H}_{13}{\bf V}_{r_{13}t_{13}3}){\bf s}_{r_{13}t_{13}}  \nonumber \\
 &+({\bf H}_{12}{\bf V}_{r_{12}t_{23}2}+{\bf H}_{13}{\bf V}_{r_{12}t_{23}3}){\bf s}_{r_{12}t_{23}}+({\bf H}_{12}{\bf V}_{r_{23}t_{23}2}+{\bf H}_{13}{\bf V}_{r_{23}t_{23}3}){\bf s}_{r_{23}t_{23}}  \nonumber \\
 &+({\bf H}_{12}{\bf V}_{r_{13}t_{23}2}+{\bf H}_{13}{\bf V}_{r_{13}t_{23}3}){\bf s}_{r_{13}t_{23}},
 \end{split}
 \end{eqnarray}
where ${\bf s}_{r_{12}t_{12}}$, ${\bf s}_{r_{13}t_{12}}$, ${\bf s}_{r_{12}t_{23}}$, ${\bf s}_{r_{13}t_{23}}$, ${\bf s}_{r_{12}t_{13}}$ and ${\bf s}_{r_{13}t_{13}}$ are desired by receiver $1$; ${\bf s}_{r_{23}t_{12}}$, ${\bf s}_{r_{23}t_{23}}$ and ${\bf s}_{r_{23}t_{13}}$ are the interference signals. The precoding matrices can be designed to satisfy :
\begin{align}
{\bf H}_{11}{\bf V}_{r_{23}t_{12}1}=-{\bf H}_{12}{\bf V}_{r_{23}t_{12}2}=[{\bf I},{\bf 0},{\bf 0},{\bf 0},{\bf 0},{\bf 0}]^{T}\nonumber \\
{\bf H}_{11}{\bf V}_{r_{23}t_{13}1}=-{\bf H}_{13}{\bf V}_{r_{23}t_{13}3}=[{\bf 0},{\bf I},{\bf 0},{\bf 0},{\bf 0},{\bf 0}]^{T} \\
{\bf H}_{12}{\bf V}_{r_{23}t_{23}2}=-{\bf H}_{13}{\bf V}_{r_{23}t_{23}3}=[{\bf 0},{\bf 0},{\bf I},{\bf 0},{\bf 0},{\bf 0}]^{T} \nonumber
\end{align}
where ${\bf I}$ and ${\bf 0}$ denote the $\frac{M}{6}\times\frac{M}{6}$ identity matrix and the $\frac{M}{6}\times\frac{M}{6}$ zero matrix, respectively. The equations for receiver $2$ and $3$
can be obtained similarly. In this way, all the interferences are neutralized and the desired signals can be successfully decoded at each
receiver by using a zero-forcing matrix. Thus, $M$ DoF per user can be obtained.

\emph{(2) the achievability of $Q_2$}:
We intend to achieve DoF per user $d=2M$ at antenna configuration $N=3M$. In this case, we use the alternating schemes as shown in Fig. \ref{ala12}. In each phase, each signal is sent simultaneously from two transmitters. If we let different transmitters send different signals, we can use the same transmission scheme as the one in the $3\times 3$ MIMO X-multicast channel at point $Q_2$ in the proof of Lemma~\ref{X-multicast}.  So the DoF per user is equal to $2M$.

\begin{figure*}[t]
\begin{centering}
\includegraphics[scale=0.7]{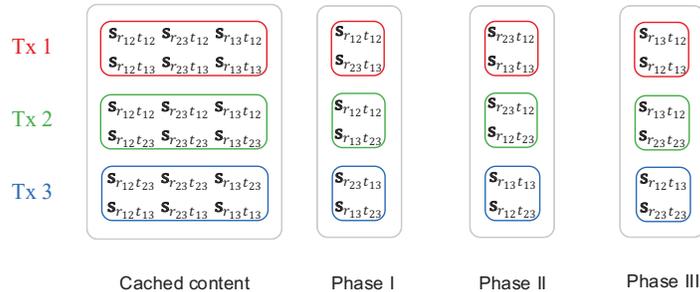}
 \caption{The alternating transmission scheme in the $3\times 3$ partially cooperative MIMO  X-multicast channel.}\label{ala12}
\end{centering}
\end{figure*}

\section*{Appendix E: proof of Lemma~\ref{fcX-multicast}}
The DoF per user of the $3\times 3$ fully cooperative MIMO X-multicast channel, shown in \eqref{dofa13} of Lemma~\ref{fcX-multicast}, is illustrated in Fig. \ref{dtotal}(d). We only need to show that the DoF per user $d=2M$ is achievable at antenna configuration $N=2M$, i.e., point $Q_1$ in Fig. \ref{dtotal}(d).

Let ${\bf s}_{r_{jk}}$ be the  $M\times 1$ transmitted signal vector for $W^{\oplus}_{r_{jk}t_{123}}$, intended to receive multicast group $\{j,k\}$ from the transmit cooperation group $\{1,2,3\}$. The received signal at receiver $j$ can be expressed as (ignoring the noise for brevity)
\begin{equation}
{\bf y}_j={\bf H}_j({\bf V}_{r_{12}}{\bf s}_{r_{12}}+{\bf V}_{r_{13}}{\bf s}_{r_{13}}+{\bf V}_{r_{23}}{\bf s}_{r_{23}}),
\end{equation}
where ${\bf H}_j=[{\bf H}_{j1},{\bf H}_{j2},{\bf H}_{j3}]$ is the $2M\times 3M$ channel matrix from the transmit cooperation group to receiver $j$, and ${\bf V}_{r_{jk}}$ is the  $3M\times M$ precoding matrix of signal ${\bf s}_{r_{jk}}$. The precoding matrices can be designed as
\begin{equation}
{\bf V}_{r_{23}}\subseteq \text{null}\left[{\bf H}_{1}\right],\quad {\bf V}_{r_{13}}\subseteq\text{null}\left[{\bf H}_{2}\right], \quad {\bf V}_{r_{12}}\subseteq \text{null}\left[{\bf H}_{3}\right].
\end{equation}
In this way, each receiver can decode its two desired signals and $2M$ DoF per user can be achieved.

\section*{Appendix F: proof of Lemma~\ref{entropy_signal}}
This is an extension of \cite[Lemma 1]{simeone} to MIMO case. The differential entropy of the received signals from any $l$ antennas can be upper bounded by
\begin{subequations}
\begin{align}
h\bigg({\bf y}_{[1:l]}^T\bigg)&\leq\sum\limits_{\iota=1}^{l}\sum\limits_{t=1}^{T}h\bigg(\sum\limits_{m=1}^{3M}h_{\iota m}x_m(t)+n_{\iota}(t)\bigg)\\
&\leq\sum\limits_{\iota=1}^{l}\sum\limits_{t=1}^{T}\log\bigg(2\pi e \text {Var}\bigg[\sum\limits_{m=1}^{3M}h_{\iota m}x_m(t)+n_{\iota}(t)\bigg]\bigg) \\ \label{47d}
&=\sum\limits_{\iota=1}^{l}\sum\limits_{t=1}^{T}\log\bigg(2\pi e\bigg(\text {Var}\bigg[\sum\limits_{m=1}^{3M}h_{\iota m}x_m(t)\bigg]+1\bigg)\bigg) \\ \label{47e}
&\leq\sum\limits_{\iota=1}^{l}\sum\limits_{t=1}^{T} \log\bigg(2\pi e \bigg(\sum\limits_{m=1}^{3M}h_{\iota m}^{2} \text {Var}\big[x_m(t)\big]+\sum\limits_{m\neq n} h_{\iota m}h_{\iota n} \sqrt{\text {Var} \big[x_m(t)\big]\text {Var} \big[x_{n}(t)\big]}+1\bigg)\bigg) \\
& \leq\sum\limits_{\iota=1}^{l}\sum\limits_{t=1}^{T} \log\bigg(2\pi e \bigg(\sum\limits_{m=1}^{3M}h_{\iota m}^{2} \text {Var}\big[x_m(t)\big]+\sum\limits_{m\neq n} h_{\iota m}h_{\iota n} \frac{\text{Var}\big[x_m(t)\big]+\text {Var} \big[x_{n}(t)\big]}{2}+1\bigg)\bigg)  \\ \label{47g}
&\leq\sum\limits_{\iota=1}^{l}\sum\limits_{t=1}^{T} \log\bigg(2\pi e\bigg(\tilde{c}P\frac{3(M+1)}{2}+1\bigg)\bigg) \\
&\leq lT\log\bigg(2\pi e\bigg(cP+1\bigg)\bigg)
\end{align}
\end{subequations}
where \eqref{47d} comes from the fact that the noise is uncorrelated with transmitted signals and is i.i.d; \eqref{47e} comes from Cauchy-Schwartz Inequality; $\tilde{c}$ is defined as $\tilde{c}=\max\limits_{m} h_{\iota m}^2$; $c$ is defined as $\frac{3\tilde{c}(M+1)}{2}$.

\end{small}

\bibliographystyle{IEEEtran}
\bibliography{IEEEabrv,reference}

\begin{thebibliography}{10}
\providecommand{\url}[1]{#1}
\csname url@samestyle\endcsname
\providecommand{\newblock}{\relax}
\providecommand{\bibinfo}[2]{#2}
\providecommand{\BIBentrySTDinterwordspacing}{\spaceskip=0pt\relax}
\providecommand{\BIBentryALTinterwordstretchfactor}{4}
\providecommand{\BIBentryALTinterwordspacing}{\spaceskip=\fontdimen2\font plus
\BIBentryALTinterwordstretchfactor\fontdimen3\font minus
  \fontdimen4\font\relax}
\providecommand{\BIBforeignlanguage}[2]{{%
\expandafter\ifx\csname l@#1\endcsname\relax
\typeout{** WARNING: IEEEtran.bst: No hyphenation pattern has been}%
\typeout{** loaded for the language `#1'. Using the pattern for}%
\typeout{** the default language instead.}%
\else
\language=\csname l@#1\endcsname
\fi
#2}}
\providecommand{\BIBdecl}{\relax}
\BIBdecl

\bibitem{long}
Y.~Cao, F.~Xu, K.~Liu, and M.~Tao, ``A storage-latency tradeoff study for
  cache-aided {MIMO} interference networks,'' in \emph{Proc. IEEE Globecom},
  Dec. 2016, pp. 1--6.

\bibitem{cisco}
Cisco, ``Cisco visual networking index: Global mobile data traffic forecast
  update 2015-2020,'' \emph{White Paper}, Feb. 2016.

\bibitem{caching1}
G.~Paschos, E.~Bastug, I.~Land, G.~Caire, and M.~Debbah, ``Wireless caching:
  technical misconceptions and business barriers,'' \emph{IEEE Commun. Mag.},
  vol.~54, no.~8, pp. 16--22, Aug. 2016.

\bibitem{caching2}
H.~Liu, Z.~Chen, and L.~Qian, ``The three primary colors of mobile systems,''
  \emph{IEEE Commun. Mag.}, vol.~54, no.~9, pp. 15--21, Sep. 2016.

\bibitem{caching3}
M.~A. Maddah-Ali and U.~Niesen, ``Coding for caching: fundamental limits and
  practical challenges,'' \emph{IEEE Commun. Mag.}, vol.~54, no.~8, pp. 23--29,
  Aug. 2016.

\bibitem{dowdy1982}
L.~W. Dowdy and D.~V. Foster, ``Comparative models of the file assignment
  problem,'' \emph{ACM Comput. Surv.}, vol.~14, no.~2, pp. 287--313, Jun. 1982.

\bibitem{zhang2015fundamental}
J.~Zhang and P.~Elia, ``Fundamental limits of cache-aided wireless {BC}:
  Interplay of coded-caching and csit feedback,'' \emph{IEEE Trans. Inf.
  Theory}, vol.~63, no.~5, pp. 3142--3160, May 2017.

\bibitem{Ali2}
Maddah-Ali and U.~Niesen, ``Cache-aided interference channels,'' in \emph{Proc.
  IEEE ISIT}, Jun. 2015, pp. 809--813.

\bibitem{xu}
F.~Xu, M.~Tao, and K.~Liu, ``Fundamental tradeoff between storage and latency
  in cache-aided wireless interference networks,'' \emph{arXiv preprint
  arXiv:1605.00203}, 2016.

\bibitem{xu2}
F.~Xu, K.~Liu, and M.~Tao, ``Cooperative {Tx/Rx} caching in interference
  channels: A storage-latency tradeoff study,'' in \emph{Proc. IEEE ISIT}, Jul.
  2016, pp. 2034--2038.

\bibitem{niesendof}
J.~Hachem, U.~Niesen, and S.~Diggavi, ``Degrees of freedom of cache-aided
  wireless interference networks,'' \emph{arXiv preprint arXiv:1606.03175},
  2016.

\bibitem{Ali3}
N.~Naderializadeh, M.~A. Maddah-Ali, and A.~S. Avestimehr, ``Fundamental limits
  of cache-aided interference management,'' \emph{IEEE Trans. Inf. Theory},
  vol.~63, no.~5, pp. 3092--3107, May 2017.

\bibitem{Xinping_topo}
X.~Yi and G.~Caire, ``Topological coded caching,'' in \emph{Proc. IEEE ISIT},
  Jul. 2016, pp. 2039--2043.

\bibitem{d2d_1}
M.~Ji, G.~Caire, and A.~F. Molisch, ``Fundamental limits of caching in wireless
  {D2D} networks,'' \emph{IEEE Trans. Inf. Theory}, vol.~62, no.~2, pp.
  849--869, Feb. 2016.

\bibitem{d2d_twc1}
N.~Golrezaei, P.~Mansourifard, A.~F. Molisch, and A.~G. Dimakis, ``Base-station
  assisted device-to-device communications for high-throughput wireless video
  networks,'' \emph{IEEE Trans. Wireless Commun.}, vol.~13, no.~7, pp.
  3665--3676, Jul. 2014.

\bibitem{d2d_twc2}
Y.~Shen, C.~Jiang, T.~Q.~S. Quek, and Y.~Ren, ``Device-to-device-assisted
  communications in cellular networks: An energy efficient approach in downlink
  video sharing scenario,'' \emph{IEEE Trans. Wireless Commun.}, vol.~15,
  no.~2, pp. 1575--1587, Feb. 2016.

\bibitem{fogRan}
A.~Sengupta, R.~Tandon, and O.~Simeone, ``Cloud and cache-aided wireless
  networks: Fundamental latency trade-offs,'' \emph{arXiv preprint
  arXiv:1605.01690}, 2016.

\bibitem{Alilimits}
M.~A. Maddah-Ali and U.~Niesen, ``Fundamental limits of caching,'' \emph{IEEE
  Trans. Inf. Theory}, vol.~60, no.~5, pp. 2856--2867, May 2014.

\bibitem{2use_ic}
S.~A. Jafar and M.~J. Fakhereddin, ``Degrees of freedom for the {MIMO}
  interference channel,'' \emph{IEEE Trans. Inf. Theory}, vol.~53, no.~7, pp.
  2637--2642, Jul. 2007.

\bibitem{Cadambe}
V.~R. Cadambe and S.~A. Jafar, ``Interference alignment and degrees of freedom
  of the {K}-user interference channel,'' \emph{IEEE Trans. Inf. Theory},
  vol.~54, no.~8, pp. 3425--2441, Aug. 2008.

\bibitem{wang2014subspace}
C.~Wang, T.~Gou, and S.~A. Jafar, ``Subspace alignment chains and the degrees
  of freedom of the three-user {MIMO} interference channel,'' \emph{IEEE Trans.
  Inf. Theory}, vol.~60, no.~5, pp. 2432--2479, May 2014.

\bibitem{ic_generalsetting}
L.~Ke, A.~Ramamoorthy, Z.~Wang, and H.~Yin, ``Degrees of freedom region for an
  interference network with general message demands,'' \emph{IEEE Trans. Inf.
  Theory}, vol.~58, no.~6, pp. 3787--3797, Jun. 2012.

\bibitem{Maddah}
M.~A. Maddah-Ali, A.~S. Motahari, and A.~K. Khandani, ``Communication over
  {MIMO} {X} channels: Interference alignment, decomposition, and performance
  analysis,'' \emph{IEEE Trans. Inf. Theory}, vol.~54, no.~8, pp. 3457--3470,
  Aug. 2008.

\bibitem{cadambe2009xchannel}
V.~R. Cadambe and S.~A. Jafar, ``Interference alignment and the degrees of
  freedom of wireless {X} networks,'' \emph{IEEE Trans. Inf. Theory}, vol.~55,
  no.~9, pp. 3893--3908, Sep. 2009.

\bibitem{sun_xchannel}
H.~Sun, T.~Gou, and S.~A. Jafar, ``Degrees of freedom of {MIMO} {X} networks:
  Spatial scale invariance and one-sided decomposability,'' \emph{IEEE Trans.
  Inf. Theory}, vol.~59, no.~12, pp. 8377--8385, Dec. 2013.

\bibitem{simeone}
A.~Sengupta, R.~Tandon, and O.~Simeone, ``Cache aided wireless networks:
  Tradeoffs between storage and latency,'' in \emph{Proc. IEEE CISS}, Mar.
  2016, pp. 320--325.

\bibitem{Wang_it16}
M.~Zamanighomi and Z.~Wang, ``Degrees of freedom region of wireless {X}
  networks based on real interference alignment,'' \emph{IEEE Trans. Inf.
  Theory}, vol.~62, no.~4, pp. 1931--1941, Apr. 2016.

\bibitem{jafar1}
S.~A. Jafar, ``Interference alignment - a new look at signal dimensions in a
  communication network,'' \emph{Found. Trends Commun. Inf. Theory}, vol.~7,
  no.~1, pp. 1--136, Jun. 2011.

\bibitem{Davidfeasibility}
G.~Bresler, D.~Cartwright, and D.~Tse, ``Feasibility of interference alignment
  for the {MIMO} interference channel,'' \emph{IEEE Trans. Inf. Theory},
  vol.~60, no.~9, pp. 5573--5586, Sep. 2014.

\bibitem{dofbroadcast}
H.~Weingarten, Y.~Steinberg, and S.~Shamai, ``The capacity region of the
  {Gaussian} multiple-input multiple-output broadcast channel,'' \emph{IEEE
  Trans. Inf. Theory}, vol.~52, no.~9, pp. 3936--3964, Sep. 2006.

\bibitem{antenna_deactivation}
R.~Wang and X.~Yuan, ``{MIMO} multiway relaying with pairwise data exchange: A
  degrees of freedom perspective,'' \emph{IEEE Trans. Signal Process.},
  vol.~62, no.~20, pp. 5294--5307, Oct. 2014.

\bibitem{cover_it}
T.~M. Cover and J.~A. Thomas, \emph{Elements of information theory}.\hskip 1em
  plus 0.5em minus 0.4em\relax John Wiley \& Sons, 2012.

\bibitem{Liu4}
K.~Liu and M.~Tao, ``Generalized signal alignment: On the achievable {DoF} for
  multi-user {MIMO} two-way relay channels,'' \emph{IEEE Trans. Inf. Theory},
  vol.~61, no.~6, pp. 3365--3386, Jun. 2015.

\end{thebibliography}

\end{document}